\documentclass[11pt, a4paper]{article}

\usepackage[numbers]{natbib}

\usepackage[utf8]{inputenc}
\usepackage[T1]{fontenc}
\usepackage{amssymb}
\usepackage{amsmath}
\usepackage{amsfonts}
\usepackage{graphicx}
\usepackage{epstopdf}
\usepackage{xcolor}
\usepackage{bm}
\usepackage[noend]{algpseudocode}
\usepackage{algorithm}
\usepackage{array}
\usepackage{multirow}
\usepackage{tikz}
\usepackage[normalem]{ulem}
\usepackage{hyperref}
\usepackage{geometry}
\usepackage{authblk}

\newcommand{\copyrightnotice}{%
\thanks{This paper has been published as an open-access article in
\textit{Energy} (Elsevier, 2025).
DOI: \url{https://doi.org/10.1016/j.energy.2025.136249}}

}

\title{Multiobjective Model Predictive Control for Residential Demand Response Management Under Uncertainty\copyrightnotice}

\author[1]{Guan-Ting Lin}
\author[2]{Wei-Yu Chiu}
\author[3]{Chien-Feng Wu}
\author[2]{Asef Nazari}
\author[2]{Dhananjay Thiruvady}

\affil[1]{Department of Electrical Engineering, National Tsing Hua University, Hsinchu, Taiwan}
\affil[2]{School of IT, Deakin University, Victoria, Australia}
\affil[3]{Department of Electrical Engineering, National Taipei University, New Taipei City, Taiwan}

\date{} 

\begin{document}

\maketitle

\begin{abstract}
Residential users in demand response programs must balance electricity costs and user dissatisfaction under real-time pricing. This study proposes a multiobjective model predictive control approach for home energy management systems with battery storage, aiming to minimize both objectives while mitigating uncertainties. Laguerre functions parameterize control signals, transforming the optimization problem into one with linear inequalities for efficient exploration. A constrained multiobjective evolutionary algorithm, incorporating convex sampler-based crossover and mutation, is developed to ensure feasible solutions. Simulations show that the proposed method outperforms existing approaches, limiting cost increases to 0.52\% under uncertainties, compared to at least 2.3\% with other methods.
\end{abstract}


\section{Introduction}\label{sec_Introduction}

Unlike many conventional commodities, electricity cannot be efficiently stored in large quantities, making the balance between generation and demand critical in power systems. The rapid growth in electricity demand, driven by population and industrial expansion, poses challenges for energy system operators, necessitating intelligent and responsive demand management. Demand response offers a cost-effective solution by enabling changes in electricity usage based on fluctuating prices \cite{pham2021deep}. Utilities and energy distribution companies encourage end-users to shift or reduce non-essential loads during peak demand periods, creating a smarter, more interactive grid. Residential users can leverage home energy management systems (HEMS) to manage smart devices and participate in demand response programs \cite{zhou2016smart}. Equipped with energy storage systems, HEMS can store electricity during low-price periods for use during high-price periods, particularly when cost minimization is a priority \cite{11Hegde}. Additionally, HEMS can adopt strategies like scheduling appliance usage \cite{15Paterakis} or integrating renewable energy sources \cite{18Melhem} to lower electricity bills. However, these strategies can impact user comfort, as shifting or curtailing energy use may lead to inconvenience. As a result, residential users often seek a balance between reducing energy costs and maintaining comfort, underscoring the need for approaches that address both objectives \cite{das2020model}.

Several studies have focused on balancing electricity costs and user comfort. Althaher et al. \cite{15Althaher} proposed an optimization-based HEMS controller to minimize electricity payments and the volume of curtailed energy, which serves as an indicator of user discomfort, using an iterative algorithm. However, combining these objectives into a single-objective problem through a linear weighted-sum approach introduces challenges. Determining appropriate weights is nontrivial, as the objectives often have different units, and their combination may lack meaningful interpretation. For instance, the weighted-sum method has been applied to combine competing objectives in a mixed-integer nonlinear programming problem for optimal scheduling of household appliances \cite{14Setlhaolo} and multiclass appliances in the smart grid \cite{16Roh}. However, the weight coefficients, typically predefined by decision-makers, are not derived systematically, and the final solution is highly sensitive to the choice of these coefficients \cite{10Pindoriya}

In contrast to single-objective models, multiobjective  models provide a Pareto front, avoiding the limitations of methods like the weighted-sum approach. multiobjective  models have been proposed for managing residential energy resources \cite{17Soares} and for domestic load scheduling \cite{14Soares}, aiming to optimize electricity costs and dissatisfaction levels concurrently using multiobjective  evolutionary algorithms (MOEAs). Additionally, Chang et al. \cite{19Chang} introduced a multiobjective  model that incorporates energy costs and privacy considerations for smart homes. Pareto optimal demand response (PODR) has been proposed to address constrained multiobjective  problems, including energy cost minimization and load factor maximization \cite{20Chiu}. Furthermore, Chiu et al. \cite{15Chiu} developed a multiobjective  immune algorithm (MOIA) to solve constrained multiobjective  problems involving three distinct objectives in a multi-microgrid system.

The choice of an MOEA over simpler metaheuristics such as particle swarm optimization (PSO) is driven by the characteristics and requirements of the multiobjective problem addressed in this study. MOEAs are particularly well-suited for solving  multiobjective problems due to their population-based nature, which allows for exploration and maintenance of diverse solutions across the Pareto front in a single run. This capability is crucial for providing a comprehensive set of trade-off solutions.
While simpler algorithms like PSO can be adapted for multiobjective optimization, they often require additional mechanisms to manage diversity and convergence, such as the use of external archives or hybrid approaches, which can increase their complexity. In contrast, MOEAs are inherently designed to address these challenges, offering a more robust and scalable framework.

Another key challenge in solving energy-related problems is the uncertainty in technical coefficients, such as renewable energy generation, electricity prices, and residential demand \cite{mavromatidis2018review}. Techniques like approximating density functions \cite{shirsat2021quantifying}, Monte Carlo simulation \cite{he2012residential}, and stochastic optimization \cite{15Nguyen} have been used to address data uncertainty, but these approaches face computational challenges. Monte Carlo methods require extensive computation due to the large number of scenarios, density approximation demands substantial data, and stochastic optimization can be numerically intractable. To overcome these limitations, several studies have adopted strategies that avoid forecasting errors in models. Rolling horizons have been employed to continuously update predictions and reduce uncertainty \cite{19Li}, while Model Predictive Control (MPC) has been applied to enhance adaptability \cite{19Luo}. Real-time electricity scheduling with rolling horizons has also been proposed to address prediction errors in renewable energy and loads \cite{16Zhang}.  For a comprehensive study of MPC-based residential demand management, refer to \cite{farrokhifar2021model}.

Recently, the concepts of MPCs and demand response have been further explored using multiobjective models to optimize residential energy management systems. P. Hua et al. \cite{hua2024multi} proposed a multiobjective MPC (MOMPC) to enhance residential energy efficiency and consumer comfort, achieving over a 16\% improvement compared to proportional-integral-derivative controls. A fuzzy logic system for managing the temperature of a multi-room residential building in response to dynamic pricing was developed in \cite{hua2024integrated}. Additionally, Jin et al. \cite{jin2017user} presented a multiobjective model to assess and incorporate user preferences, while Freire et al. \cite{freire2020optimal} focused on maximizing the economic benefits of microgrids and minimizing energy storage system degradation in the presence of renewable resources. These studies primarily utilize weighted-sum methods to address multiple objectives, which, while straightforward, are constrained by the limitations discussed earlier.

To take advantage of both MOEAs and MPC, 
a few studies have investigated MOMPC recently; it extends the traditional MPC framework by explicitly incorporating multiobjective optimization methods to address tradeoffs between objectives, rather than focusing on a single cost function. For instance, 
Yao et al. \cite{yao2023multi} developed a multi-level and multiobjective optimization model for real-time operation of integrated energy systems, addressing renewable generation uncertainty using conditional value at risk theory and improving operational strategies with MPC.
Ascione et al. \cite{ascione2016simulation} developed a simulation-based MPC approach that uses multiobjective  optimization to balance operating costs and thermal comfort in building energy management, achieving significant cost reductions and comfort improvements compared to standard control strategies.
  
Existing MOMPC methods are typically based on the MOEA NSGA-II and employ constraint-domination concepts  \cite{12Salinas} or penalty methods \cite{09Pedrasa} to handle physical constraints. However, these methods often become inefficient when dealing with dynamic constraints like energy storage, as the random generation of power control signals over time horizons frequently violates storage constraints. This results in excessive computation on infeasible solutions and potentially degrades solution quality, underscoring the need for a more efficient search method specifically designed for energy storage dynamics.

To address these challenges, this study proposes a novel MOMPC framework for HEMS that simultaneously optimizes energy cost and user dissatisfaction while considering the constraints of power-flexible and time-flexible appliances and energy storage system dynamics.
Three sources of uncertainty are considered: Power demand, renewable power generation, and electricity prices. 
Laguerre functions are introduced to parameterize control signals, transforming the optimization problem into one governed by linear inequalities and facilitating efficient exploration of the feasible region. Additionally, a convex sampler is integrated into the MOEA framework to focus the search within the feasible set, ensuring valid solutions, reducing computational waste on infeasible points, and significantly accelerating convergence. This approach addresses the inefficiencies of exploring infeasible regions and enhances the overall performance of the MOMPC method. 

Table~\ref{tab:comparison} summarizes relevant approaches to  residential demand response. The comparative analysis of existing home energy management approaches highlights distinct tradeoffs in computational efficiency, uncertainty handling, and optimization effectiveness. The weighted-sum approach, despite its simplicity, struggles with determining appropriate weights and capturing Pareto optimal solutions. MOEAs offer a comprehensive Pareto front but suffer from excessive computation on infeasible solutions. Stochastic approaches explicitly model uncertainties but often require high computational resources and are limited to single-objective formulations. Rolling horizon and MPC methods dynamically adapt to changing conditions but traditionally focus on single-objective optimization, limiting their ability to balance competing priorities effectively. While existing MOMPC improves upon conventional MPC by integrating multiobjective optimization, it still inherits inefficiencies in constraint handling and solution feasibility.
The proposed MOMPC framework addresses these limitations by efficiently balancing energy cost and user satisfaction while ensuring solution feasibility through Laguerre function parameterization and convex sampling.

\begin{table}[htbp]
\centering
\caption{Existing Approaches for Home Energy Management Systems (HEMS)}
\label{tab:comparison}
\resizebox{\textwidth}{!}{%
\begin{tabular}{|p{2.5cm}|p{2.8cm}|p{5.5cm}|p{5.5cm}|}
\hline
\textbf{Approach} & \textbf{References} & \textbf{Pros} & \textbf{Cons} \\
\hline
{\small Weighted Sum Approach} & \cite{15Althaher}, \cite{14Setlhaolo}, \cite{16Roh} & 
{\small Simple implementation; Computationally efficient.} & 
{\small Determining appropriate weights is nontrivial; Different units for objectives; Combinations lack meaningful interpretation; High sensitivity to coefficient choice \cite{10Pindoriya}; Cannot capture all Pareto optimal solutions; Poor handling of uncertainties.} \\
\hline
{\small Multiobjective Evolutionary Algorithms (MOEAs)} & \cite{17Soares}, \cite{14Soares}, \cite{19Chang}, \cite{20Chiu}, \cite{15Chiu} & 
{\small Provides complete Pareto fronts; Captures tradeoffs between objectives; Handles multiple objectives simultaneously; Population-based approach maintains diverse solutions; Moderate ability to handle uncertainty through population diversity.} & 
{\small Difficulty handling dynamic constraints efficiently; Random generation of solutions often violates constraints; Excessive computation on infeasible solutions; Requires additional mechanisms for comprehensive uncertainty management.} \\
\hline
{\small Stochastic Approaches} & \cite{shirsat2021quantifying}, \cite{he2012residential}, \cite{15Nguyen} & 
{\small Explicit representation of uncertainty; Robust solutions; Strong theoretical foundation for uncertainty management.} & 
{\small Monte Carlo requires extensive computation; Density approximation demands substantial data; Stochastic optimization can be numerically intractable; Often limited to single-objective formulations.} \\
\hline
{\small Rolling Horizon and MPC} & \cite{19Li}, \cite{19Luo}, \cite{16Zhang}, \cite{farrokhifar2021model} & 
{\small Continuously updates predictions; Reduces impact of uncertainty; Adaptive to changing conditions; Real-time implementation possible; Strong dynamic uncertainty handling.} & 
{\small Single-objective focus in traditional implementations; Cannot capture all Pareto
optimal solutions.} \\
\hline
{\small MOMPC} & \cite{hua2024multi}, \cite{hua2024integrated}, \cite{jin2017user}, \cite{freire2020optimal}, \cite{yao2023multi}, \cite{ascione2016simulation} & 
{\small Combines benefits of MPC and multiobjective optimization; Balances multiple competing objectives; Demonstrated improvements in energy efficiency and comfort; Strong uncertainty management through prediction updates.} & 
{\small Many still rely on weighted-sum methods with their inherent limitations; Inefficient handling of dynamic constraints; Excessive computation on infeasible solutions.} \\
\hline
{\small Proposed MOMPC Framework} & {\small This study} & 
{\small Efficiently balances energy cost and user dissatisfaction; Captures Pareto optimal solutions; Strong uncertainty management;
 Parameterization via Laguerre functions facilitates efficient handling of
dynamic constraints; Convex sampler ensures feasible solutions, improving solution quality.} & 
{\small Computational overhead from Laguerre function parameterization and convex sampling.} \\  
\hline

\end{tabular}%
}
\end{table}

The key contributions of this work include leveraging multiobjective optimization while overcoming the limitations of weighted sum methods and mitigating the impact of uncertainties. The use of Laguerre functions to approximate power control signals, coupled with convex sampling techniques, enables effective handling of constrained multiobjective problems. By approximating the complex search space using linear inequalities, the proposed methodology avoids the reliance on slow-converging metaheuristics and ensures feasibility throughout the search process. These innovations collectively enhance solution quality and efficiency compared to existing approaches, making the MOMPC framework a robust tool for optimizing energy cost and user satisfaction in HEMS.

This paper is organized as follows. Section~\ref{sec_SystemModels} presents the system model for home energy management. Section~\ref{sec_Proposedapproach} details the proposed approach. Section~\ref{sec_Simulation} presents a comparative analysis of different methods through simulation results. Finally, Section~\ref{sec_Conclusion} concludes the paper.

\section{Residential System Models and Problem Formulation}\label{sec_SystemModels}

This section describes a mathematical model for an HEMS and its key components. These include home appliances, renewable energy sources like solar panels, and energy storage devices like batteries. Residential users utilize a HEMS to schedule the operation of three appliance types. To optimize energy use and cost, a HEMS considers various energy sources including renewable energy generated on-site and energy exchanged with the utility grid. Energy storage devices play a crucial role in a HEMS by enabling residents to store excess renewable energy for later use or to adjust their energy consumption based on real-time pricing schemes. Residential users consider both the energy cost and user dissatisfaction and reduce the uncertainty influence, such as power load, electricity price, and renewable generation in a HEMS simultaneously.

\subsection{Home Appliances}\label{subsec_HomeAppliances}

Home appliances can be broadly classified into three categories based on their flexibility in terms of scheduling and power consumption \cite{17Manzoor}, namely, inflexible, time-flexible, and power-flexible appliances.
This categorization reflects the diversity of household appliances typically found in residential energy systems. Inflexible appliances (e.g., refrigerators) operate continuously and cannot be easily controlled, while time-flexible appliances (e.g., washing machines) can defer operation to non-peak hours, and power-flexible appliances (e.g., HVAC systems) can adjust their energy consumption dynamically. Including these types ensures the model captures the full range of real-world appliance behaviors, making the results more applicable and realistic.

Inflexible appliances consume a fixed amount of power, denoted by $\gamma_{a}$, whenever they are turned on. Their power consumption in any time slot $t$ is simply expressed as
\begin{equation}\label{eq_Pa_t}
     P_{a}(t)=\gamma_{a}\quad \forall a\in\mathcal{A_{\text{if}}},
\end{equation}
where $\mathcal{A_{\text{if}}}$ is the set of the inflexible appliances. When a user turns on inflexible appliances between starting time $\alpha_{a}$ and ending time $\beta_{a}$, the power consumption of this type of appliances $P^{\text{con}}_{a}(t)$ is denoted as
\begin{equation}\label{eq_Pa_t_con}
   P^{\text{con}}_{a}(t)=\sum_{a\in \mathcal{A_{\text{if}}}}P_{a}(t)X_{a}(t),
   \end{equation}
where the indicator function $X_{a}(t)$ is defined as
\begin{equation}\label{eq_Xa_t}
    X_{a}(t)=
    \begin{cases}
          1, & \text{if } t\in[\alpha_{a},\beta_{a}] \\
          0, & \text{otherwise}
    \end{cases}\quad
    \forall a\in\mathcal{A_{\text{if}}}.
\end{equation}


For a time-flexible appliance $b\in\mathcal{A_{\text{tf}}}$ with constant power consumption, users have the advantage of delaying their start time (denoted as $t_{b}$) to potentially reduce energy costs.
\begin{equation}\label{eq_Pb_t}
     P_{b}(t)=\gamma_{b}\quad \forall b\in\mathcal{A_{\text{tf}}}
\end{equation}
where $\gamma_{b}$ is the amount of power consumption and $\mathcal{A_{\text{tf}}}$ represents the set of all time-flexible appliances. 
For appliance $b\in \mathcal{A_{\text{tf}}}$,
the indicator function $X_{b}(t)$ is defined as
\begin{equation}\label{eq_Xb_t}
        X_{b}(t)=
        \begin{cases}
          1, & t\in \{t_{b},\cdots,t_{b}+T_{b}-1\}\subset[\alpha_{b},\beta_{b}]\\
          0, & \text{otherwise}
        \end{cases}
\end{equation}
where $T_b$ represents the duration of
the whole operation procedure,
and $\alpha_{b}$ and $\beta_{b}$ represent the start and end of the reduced energy cost period for the time-flexible appliance, respectively.

The power consumption of a time-flexible appliance $b$ can be expressed by
\begin{equation}\label{eq_Pb_t_con}
   P^{\text{con}}_{b}(t)=\sum_{b\in \mathcal{A_{\text{tf}}}}P_{b}(t)X_{b}(t).
\end{equation}

Power-flexible appliances offer residential users the highest control, allowing them to adjust both output level and power consumption based on their needs. These appliances typically have variable power consumption, denoted as $P_{c}(t)$, which can be adjusted within a range defined by a minimum power, $\gamma_{c}^{\text{min}}$, and a maximum power, $\gamma_{c}^{\text{min}}$, i.e.,
\begin{equation}\label{eq_Pc_t}
     \gamma_{c}^{\text{min}} \leq P_{c}(t)\leq \gamma_{c}^{\text{max}} \quad
     \forall c\in\mathcal{A_{\text{pf}}}
\end{equation}
where $\mathcal{A_{\text{pf}}}$ represents the set of all power-flexible appliances. Consider $\alpha_{c}$ and $\beta_{c}$ denote the start and end time of using power-flexible appliance $c$. Its power consumption can be expressed as
\begin{equation}\label{eq_Pc_t_con}
   P^{\text{con}}_{c}(t)=\sum_{c\in \mathcal{A_{\text{pf}}}}P_{c}(t)X_{c}(t)
\end{equation}
where the indicator function $X_{c}(t)$ is defined as
\begin{equation}\label{eq_Xc_t}
   X_{c}(t)=
        \begin{cases}
          1, & \text{if } t\in[\alpha_{c},\beta_{c}] \\
          0, & \text{otherwise}
        \end{cases}\quad
        \forall c\in \mathcal{A_{\text{pf}}}.
\end{equation}

The total power consumption of home appliances $a\in \mathcal{A_{\text{if}}}, b\in\mathcal{A_{\text{tf}}}$, and $c\in \mathcal{A_{\text{pf}}}$ in time slot $t$ is expressed as
\begin{equation}\label{eq_P_t_con}
    P^{\text{con}}(t)=P^{\text{con}}_{a}(t)+P^{\text{con}}_{b}(t)+P^{\text{con}}_{c}(t).
\end{equation}

Conventional appliance scheduling methods typically rely on the assumption of perfectly predictable future load. This allows them to use any available deterministic optimization technique. By contrast, this study assumes the inherent uncertainty in future load $P^{\text{con}}(t+1), P^{\text{con}}(t+2), \ldots, P^{\text{con}}(t+M-1)$ which can only be predicted at the current time slot $t$ with some associated prediction error bounds, where $M$ represents the prediction window. 
This assumption is valid because future power demand is influenced by a variety of unpredictable factors, such as user behavior, weather conditions, and unexpected appliance usage patterns. Even though historical data can provide some insight into future demand, it is not possible to predict the exact future load with absolute certainty.

\subsection{Energy Storage System }\label{subsec_EnergyStorageDevice}

In a smart home, an energy storage device significantly enhances the flexibility of a HEMS. Renewable energy sources, like solar and wind power, offer cleaner and often cheaper alternatives to traditional fossil-based energy sources. However, their intermittent nature, meaning they are not always available when they are needed, can be a challenge. Integrating renewable energy with energy storage devices in a HEMS is a remedy to address this challenge.

This paper explores a HEMS equipped with both an energy storage device (battery pack) and renewable energy sources. The battery capacity at time $t$ is denoted as $E_{\text{s}}(t)$ and the charging/discharging power as $P_{\text{s}}(t)$. When $P_{\text{s}}(t)$ is greater than or equal to zero, $P_{\text{s}}(t)\geq0$, it represents power charging the battery. Conversely, a negative $P_{\text{s}}(t)$ value ($P_{\text{s}}(t)<0$) indicates power discharging from the battery.
For battery $s$,
the storage dynamics for the battery capacity can be expressed as
\begin{equation}\label{eq_Fel_tadd1}
E_{\text{s}}(t+1) =\rho E_{\text{s}}(t)+P_{\text{s}}(t)\Delta t
\end{equation}
where $\rho$ represents the leakage rate of the battery, and $\Delta t$ represents the duration of a time slot.

The capacity of each battery is subject to constraints, with both a lower bound and an upper bound. Additionally, limitations exist on the maximum rates for charging and discharging:
\begin{equation}\label{eq_Es_constrain}
|P_{\text{s}}(t)| \leq S^{\text{max}}
\end{equation}
\begin{equation}\label{eq_Fel_constrain}
E_{\text{s}}^{\text{min}} \leq E_{\text{s}}(t) \leq E_{\text{s}}^{\text{max}}
\end{equation}
where $E_{\text{s}}^{\text{min}}$ and $E_{\text{s}}^{\text{max}}$ respectively represent the minimum and maximum capacity, and $S^{\text{max}}$ is the maximum charging/discharging rate.

\subsection{Residential Demand Response Management Problem}\label{subsec_Optimization}
With the total appliance consumption of $P^{\text{con}}(t)\text{ kW}$, renewable energy power generation of $P_{\text{re}}(t)\text{ kW}$, and the amount of power from the battery pack $P_{\text{s}}(t)\text{ kW}$, the total power consumption can be expressed as
\begin{equation}\label{eq_Pgrid_t}
P_{\text{total}}(t)=P^{\text{con}}(t)+P_{\text{s}}(t)-P_{\text{re}}(t).
\end{equation}
To minimise energy costs, the HEMS  prioritises using renewable energy whenever available. If the total power generated by renewable sources exceeds the current demand (i.e., $P^{\text{con}}(t)+P_{\text{s}}(t)-P_{\text{re}}(t)<0$ ), the excess energy cannot be stored and will be sold back to the grid. 
 Due to the inherent variability of renewable energy sources, future power generation, $P_{\text{re}}(t+1), \ldots, P_{\text{re}}(t+M-1)$, can only be estimated at the current time slot $t$.

Two primary competing objectives are considered: minimising the electricity bill and mitigating user dissatisfaction. The total energy cost $F_{\text{cost}}$ can be calculated as\cite{20Chiu}
\begin{equation}\label{eq_Fcost}
            F_{\text{cost}}=\sum_{m=0}^{M-1} \lambda(t+m)P_{\text{total}}(t+m)\Delta t, 
\end{equation}
where $\lambda(t)$ represents the market price at time slot $t$
if $P_{\text{total}}(t)>0$; otherwise, $\lambda(t)$ represents a feed-in tariff (FIT), which is lower than the market price.
FITs are assumed to be set at a constant rate per kilowatt-hour (kWh) over a defined contract period \cite{EWOSASolarFeedInTariffs}. This provides predictability for energy producers and ensures a stable income for the electricity they export to the grid. Setting the rate lower than the market price is reasonable, as FITs are typically designed to encourage self-consumption of renewable energy while offering a modest return for any excess energy sold back to the grid.

To incorporate the price uncertainty into the proposed model (\ref{eq_Fcost}), nominal market prices (predicted values) of $\lambda(t+1), \lambda(t+2), \ldots, \lambda(t+M-1)$ are assumed to be available at the current time slot $t$. The assumption is reasonable given the increasing availability of forecasting tools and smart energy management systems that provide price predictions. Access to such information allows residential users to optimize their energy usage and align it with cost-saving strategies. Including predicted prices in the model is important because it reflects a key element of decision-making in modern energy markets.

User dissatisfaction can be modeled as the difference between the adjusted power or the start time and their nominal values \cite{16Ma}. For the time-flexible appliances, user dissatisfaction caused by the delay time between the request time $t_{b}^{\text{req}}$ and the real start time $t_{b}$ can be expressed as
\begin{equation}\label{eq_Ftf}
     F_{tf}=\sum_{b\in \mathcal{A_{\text{tf}}}}\theta_{b}(t_{b}-t_{b}^{\text{req}})^{k_{b}},
\end{equation}
where  parameter $\theta_{b}$ is the discomfort weight and  coefficient $k_{b}\geq 1$ captures some operation characteristics.

For the power flexible appliances in~(\ref{eq_Pc_t}), dissatisfaction $F_{\text{pf}}$ of using appliance $c$ caused by the difference between actual power consumption $P_{\text{c}}(t)$ and normal power $P_{c}^{\text{nor}}$ can be expressed as
\begin{equation}\label{eq_Fpf}
     F_{\text{pf}}=\sum_{m=0}^{M-1}\sum_{c\in \mathcal{A_{\text{pf}}}}\xi_{c}(P_{\text{c}}(t+m)-P_{c}^{\text{nor}})^{2},
    \end{equation}
where  coefficient $\xi_{c}$ is the discomfort weight. The square relationship in (\ref{eq_Fpf})  
is called a Taguchi loss function representing a concept in quality engineering,  which has been used to quantify the cost of deviating from a desired target value \cite{Taguchi2007}. 
In practice, 
parameters such as   $\theta_{b}$, $k_{b}$, and  $\xi_{c}$ in the user models involved in (\ref{eq_Ftf}) and (\ref{eq_Fpf})can be 
tuned using, for example,  sensitivity analysis \cite{Saltelli2008} or machine learning based methods \cite{FENG2022112357}.

The dissatisfaction induced by
time-flexible and power-flexible appliances can be expressed as
\begin{equation}\label{eq_Fdissat}
        F_{\text{dissat}}=F_{\text{tf}}+F_{\text{pf}}.
\end{equation}

With the definitions (\ref{eq_Fcost})--(\ref{eq_Fdissat}), an optimal residential demand response management can be realised by solving 
\begin{equation}\label{eq_multiobjective}
    \begin{split}
     &\mathop{\min}_{\substack{
     P_{c}(t+m), t_{b}, P_{\text{s}}(t+m)\\
     m=0\ldots M-1, b\in \mathcal{A_{\text{tf}}}, c\in \mathcal{A_{\text{pf}}}
     }}\text{ }F_{\text{cost}}\\
     &\mathop{\min}_{\substack{
     P_{c}(t+m), t_{b}\\
     m=0\ldots M-1, b\in \mathcal{A_{\text{tf}}}, c\in \mathcal{A_{\text{pf}}}
     }}\text{ }F_{\text{dissat}}\\
     &\text{subject to 
     (\ref{eq_Pa_t})--(\ref{eq_Fel_constrain}).}
    \end{split}
\end{equation}
The number of decision variables involved is $\left|\mathcal{A_{\text{tf}}} \right|+(\left|\mathcal{A_{\text{pf}}} \right|+1)M$, where $A_{\text{tf}}$ and $A_{\text{pf}}$ are related to the number of appliances, and $M$ represents the 
optimization window, also 
the prediction horizon for
the future renewable power generation, electricity prices, and base power load.
The optimization problem is always feasible, with a trivial solution obtained by setting all control signals to zero. In this case, power-flexible and time-flexible appliances operate as usual without any adjustments, and the energy storage system remains inactive.

\section{Proposed multiobjective  Model Predictive Control for Residential Demand Response Management}\label{sec_Proposedapproach}
This section details the proposed MOMPC approach, which utilizes both MPC and MOEA. 
Motivated by the problem formulation in \cite{09Wang}, 
a state-space model is developed to predict future system states. 
The Laguerre functions
are then introduced to 
approximate the optimal values of decision variables 
using the Laguerre coefficients.
The problem in  (\ref{eq_multiobjective})
can then be transformed into another constrained multiobjective optimization problem, which 
allows for
a more efficient exploration of the feasible set.
Finally, an MOEA
utilizing a mutation operation based on convex sampling and a crossover operation 
is developed to solve the transformed multiobjective optimization problem, producing control signals for residential demand response management.

\subsection{Problem Transformation Based on Laguerre functions}\label{subsec_Model}
To establish an MPC framework, the original model (\ref{eq_multiobjective}) is transformed  into a state-space representation. This allows for better prediction and control of system behavior. Within this transformation, a new state variable is introduced, denoted by $C_{\text{p}}(t)$ (cumulative power), to specifically represent the cumulative power consumption of power-flexible appliances, as defined in Equation~(\ref{eq_Pc_t}).

\begin{equation}\label{eq_Tc_tadd1}
    C_{\text{p}}(t+1)=C_{\text{p}}(t)+ \sum_{c\in \mathcal{A_{\text{pf}}}}P_{c}(t).
\end{equation}
The combination of equations (\ref{eq_Fel_tadd1}) and (\ref{eq_Tc_tadd1}) into the state-space model provides the vector representation of the dynamic state space.
\begin{equation*}
\begin{bmatrix}
    E_{\text{s}}(t+1) \\
    C_{\text{p}}(t+1)
\end{bmatrix}=
\begin{bmatrix}
    \rho & 0 \\
    0  & 1
\end{bmatrix}
\begin{bmatrix}
    E_{\text{s}}(t) \\
    C_{\text{p}}(t)
\end{bmatrix}+
\begin{bmatrix}
    \Delta t & \boldsymbol{o}_{ 1 \times \ell}\\
    0  & \boldsymbol{I}_{ 1 \times \ell}
\end{bmatrix}
\begin{bmatrix}
    P_{s}(t) \\
    P_{c_{1}}(t) \\
    \vdots \\
    P_{c_{\ell}}(t)
\end{bmatrix},
\end{equation*}
where the number of the power-flexible appliances is $\ell=\left|\mathcal{A_{\text{pf}}} \right|$.

This equation is concisely represented in the following equation.
\begin{equation}
\boldsymbol{x}(t+1)=\boldsymbol{A}\boldsymbol{x}(t)+\boldsymbol{B}%
\boldsymbol{u}(t),\label{eq_x_tadd1}%
\end{equation}
where%
\[
\boldsymbol{A}=%
\begin{bmatrix}
\rho & 0\\
0 & 1
\end{bmatrix}
,\text{ }\boldsymbol{x}(t)=%
\begin{bmatrix}
x_{1}(t)\\
x_{2}(t)
\end{bmatrix}
=%
\begin{bmatrix}
E_{\text{s}}(t)\\
C_{\text{p}}(t)
\end{bmatrix}
,
\]

\[
\boldsymbol{B}=%
\begin{bmatrix}
\Delta t & \boldsymbol{o}_{1\times\ell}\\
0 & \boldsymbol{I}_{1\times\ell}%
\end{bmatrix}
,\text{and }\boldsymbol{u}(t)=%
\begin{bmatrix}
u_{\text{s}}(t)\\
u_{1}(t)\\
\vdots\\
u_{\ell}(t)
\end{bmatrix}
=%
\begin{bmatrix}
P_{s}(t)\\
P_{c_{1}}(t)\\
\vdots\\
P_{c_{\ell}}(t)
\end{bmatrix}
.
\]
In (\ref{eq_x_tadd1}),
$\boldsymbol{u}(t)$
represents 
manipulated variables adjusted by the MPC to control the system, consisting of charging/discharging power
$u_{s}(t)=P_{s}(t)$ for the battery and 
power control $u_{i}(t)=P_{c_{i}}(t)$ of the $i$th power-flexible appliance;
in  $\boldsymbol{x}(t)$,
battery energy level
$x_1(t)=E_{\text{s}}(t)$
 is the 
feedback variable  informing the controller about the system’s current state, and
 cumulative power consumption of power-flexible appliances
$x_2(t)=C_{\text{p}}(t)$
represents a controlled variable associated
with the performance objectives the MPC seeks to meet.

To better explore the feasible set while capturing the dynamics of the control signals $P_s(t)$ and   $P_c(t)$,
the Laguerre network is used
with $J$ Laguerre coefficients 
as the basis in consideration of the prediction  horizon $M$.  
 The discrete-time Laguerre network through Z-transform is formulated as follows.
\begin{equation}\label{eq_gamman}
    \Gamma_{j}(z)=\Gamma_{j-1}(z)\frac{z^{-1}-p}{1-pz^{-1}},
    \quad j=2, 3, \ldots, J,
\end{equation}
where $\Gamma_{1}(z)=\frac{\sqrt{1-p^{2}}}{1-pz^{-1}}$ is the initial term, the parameter $p\in [0,1)$ represents scaling factor defined by users. A larger value of $p$  
implies that the Laguerre functions decay to zero at a slower speed, suitable for a larger optimization window; some experimentation can be conducted to determine an appropriate value for 
$p$ \cite{Gautschi2004}.

Denote $l_{j}(m)$ as inverse z-transform of $\Gamma_{j}(z)$. Let
\begin{equation}\label{eq_gt}
    \boldsymbol{L}(m)=\begin{bmatrix}l_{1}(m) & l_{2}(m) & l_{3}(m)& ... &l_{J}(m)  \end{bmatrix}^{T}
\end{equation}
denotes the Laguerre function that satisfies the following difference equation.
\begin{equation}\label{eq_gtadd1}
    \boldsymbol{L}(m+1)=\boldsymbol{A}_{\text{la}}  \boldsymbol{L}(m)
    \quad m=0, 1, \ldots, M-1,
\end{equation}
where $\boldsymbol{L}(0)= \sqrt{1-p^{2}}
\begin{bmatrix}
    1 & -p & p^{2} & \cdots & (-p)^{J-1}
\end{bmatrix}^{T}$, and
\begin{equation*}
    \boldsymbol{A}_{\text{la}}=\begin{bmatrix}
    p & 0 & ...&0 \\
    (1-p^{2}) & p & ...&0 \\
    -p(1-p^{2}) & (1-p^{2}) & ...&0 \\
    p^{2}(1-p^{2}) & -p(1-p^{2}) & ...&0\\

    \vdots & \vdots & \ddots &\vdots\\
    -p^{J-2}(1-p^{2}) & p^{J-3}(1-p^{2}) & ...&p
    \end{bmatrix}.
\end{equation*}

The variable vector $\boldsymbol{ u}(t)$ is approximated by the orthonormal Laguerre function as depicted in the following equation.
\begin{equation}\label{eq_ut}
\begin{aligned}
   \boldsymbol{u}(t+m)
   &=\begin{bmatrix}
   u_{\text{s}}(t+m)\\
   u_{1}(t+m)\\
   \vdots\\
   u_{\ell}(t+m)
   \end{bmatrix}
   \approx\begin{bmatrix}
   \boldsymbol{ L}(m)^{T}
\boldsymbol{\eta}_{\text{s}}\\
   \boldsymbol{ L}(m)^{T}\boldsymbol{\eta}_{1}\\
   \vdots\\
   \boldsymbol{ L}(m)^{T}\boldsymbol{\eta}_{\ell}
   \end{bmatrix}\\
   &=\begin{bmatrix}
   \boldsymbol{ L}(m)^{T} & \boldsymbol{ o}_{1 \times J} & \cdots & \boldsymbol{ o}_{1 \times J} \\
   \boldsymbol{ o}_{1 \times J} & \boldsymbol{ L}(m)^{T} &  \cdots & \boldsymbol{ o}_{1 \times J} \\
   \vdots &\vdots&\ddots&\vdots\\
   \boldsymbol{ o}_{1 \times J} & \boldsymbol{ o}_{1 \times J} & \cdots &\boldsymbol{ L}(m)^{T}
   \end{bmatrix}\boldsymbol{\eta}=\boldsymbol{G}(m)\boldsymbol{\eta},
\end{aligned}
\end{equation}
where the coefficient of the Laguerre function is defined as
\begin{equation}\label{eq_eta}
    \boldsymbol{\eta}=\begin{bmatrix}
    \boldsymbol{\eta}_{\text{s}}^{T} & \boldsymbol{\eta}_{1}^{T} &\cdots&\boldsymbol{\eta}_{\ell}^{T}
   \end{bmatrix}^{T}.
\end{equation}

After reducing the number of decision variables using the Laguerre function, the state variables are expressed considering Equation~(\ref{eq_x_tadd1}). Here,  $\boldsymbol{ x}(t+m|t)$ denotes the predicted states for future time slots $t+m$ based on the information available at the current time slot, $t$. The expression for these future state variables are expressed as follows.


\begin{equation*}
\left.
\begin{array}
[c]{l}%
\boldsymbol{x}(t+1|t)=\boldsymbol{A}\boldsymbol{x}(t)+\boldsymbol{Bu}(t)\\
\boldsymbol{x}(t+2|t)=\boldsymbol{A}\boldsymbol{x}(t+1)+\boldsymbol{Bu}(t+1)\\
\text{ \ \ \ \ \ \ \ \ \ \ \ \ \ }=\boldsymbol{A}^{2}\boldsymbol{x}%
(t)+\boldsymbol{ABu}(t)+\boldsymbol{Bu}(t+1)\\
\boldsymbol{x}(t+3|t)=\boldsymbol{A}\boldsymbol{x}(t+2)+\boldsymbol{Bu}(t+2)\\
\text{ \ \ \ \ \ \ \ \ \ \ \ \ \ }=\boldsymbol{A}^{3}\boldsymbol{x}%
(t)+\boldsymbol{A}^{2}\boldsymbol{Bu}(t)+\boldsymbol{A}\boldsymbol{Bu}%
(t+1)+\boldsymbol{Bu}(t+2)\\
\text{ \ \ \ \ \ \ \ \ \ \ \ \ \ \ \ \ \ \ \ \ \ \ \ \ \ \ \ \ \ \ \ \ \ \ }%
\vdots\\
\boldsymbol{x}(t+m|t)=\boldsymbol{A}\boldsymbol{x}(t+m-1)+\boldsymbol{Bu}%
(t+m-1)\\
\text{ \ \ \ \ \ \ \ \ \ \ \ \ \ \ }=\boldsymbol{A}^{m}\boldsymbol{x}%
(t)+\sum_{i=0}^{m-1}\boldsymbol{A}^{m-i-1}\boldsymbol{Bu}(t+i)\\
\text{ \ \ \ \ \ \ \ \ \ \ \ \ \ \ }=\boldsymbol{A}^{m}\boldsymbol{x}%
(t)+\boldsymbol{\varphi}(m)\boldsymbol{\eta},
\end{array}
\right.  %
\end{equation*}
where
\begin{equation*}
\boldsymbol{\varphi}(m)=\sum_{i=0}^{m-1}\mathbf{A}^{m-i-1}\mathbf{B}\boldsymbol{G}(m).
\end{equation*}

The next step is to 
transform the constraints (\ref{eq_Pc_t}), (\ref{eq_Es_constrain}), and (\ref{eq_Fel_constrain}) 
into linear inequalities 
by approximating the control signals using the Laguerre functions. The power-flexible appliances constraints in~(\ref{eq_Pc_t}) is converted into
\begin{equation*}
     \gamma_{c}^{\text{min}} \leq P_{c}(t)\leq \gamma_{c}^{\text{min}} \quad
     \forall c\in\mathcal{A_{\text{pf}}},
\end{equation*}
 and the charging/discharging power constraint (\ref{eq_Es_constrain}) is rewritten as 
\begin{equation*}
     -S^{\text{max}} \leq P_{\text{s}}(t) \leq S^{\text{max}}
\end{equation*} with the vector representation of 
\begin{equation}\label{eq_lagu_combine}
  \begin{bmatrix}
    -S^{\text{max}}\\
    \gamma_{c_{1}}^{\text{min}}\\
    \vdots\\
    \gamma_{c_{\ell}}^{\text{min}}
   \end{bmatrix}
   \leq
\begin{bmatrix}
    P_{s}(t) \\
    P_{c_{1}}(t) \\
    \vdots \\
    P_{c_{\ell}}(t)
  \end{bmatrix}
  \leq
  \begin{bmatrix}
    S^{\text{max}}\\
    \gamma_{c_{1}}^{\text{max}}\\
    \vdots\\
    \gamma_{c_{\ell}}^{\text{max}}
   \end{bmatrix}.
\end{equation}
Inequality (\ref{eq_lagu_combine}), considering equation (\ref{eq_ut}), can be written as the following linear inequalities.
\begin{equation}\label{eq_lagu_umax}
\begin{bmatrix}
   \boldsymbol{ L}(m)^{T} & \boldsymbol{o}_{1 \times J} & \cdots & \boldsymbol{ o}_{1 \times J} \\
    \boldsymbol{o}_{1 \times J} & \boldsymbol{ L}(m)^{T} &  \cdots & \boldsymbol{ o}_{1 \times J} \\
   \vdots &\vdots&\ddots&\vdots\\
   \boldsymbol{ o}_{1 \times J} & \boldsymbol{ o}_{1 \times J} & \cdots &\boldsymbol{ L}(m)^{T}
   \end{bmatrix}\boldsymbol{\eta}
   \leq
   \begin{bmatrix}
    S^{\text{max}}\\
    \gamma_{c_{1}}^{\text{max}}\\
    \vdots\\
    \gamma_{c_{\ell}}^{\text{max}}
   \end{bmatrix},
\end{equation} and
\begin{equation}\label{eq_lagu_umin}
   -\begin{bmatrix}
   \boldsymbol{ L}(m)^{T} & \boldsymbol{ o}_{1 \times J} & \cdots & \boldsymbol{ o}_{1 \times J} \\
    \boldsymbol{o}_{1 \times J} & \boldsymbol{ L}(m)^{T} &  \cdots & \boldsymbol{ o}_{1 \times J} \\
   \vdots &\vdots&\ddots&\vdots\\
   \boldsymbol{ o}_{1 \times J} & \boldsymbol{ o}_{1 \times J} & \cdots &\boldsymbol{ L}(m)^{T}
   \end{bmatrix}\boldsymbol{\eta}
   \leq
   -\begin{bmatrix}
    -S^{\text{max}}\\
    \gamma_{c_{1}}^{\text{min}}\\
    \vdots\\
    \gamma_{c_{\ell}}^{\text{min}}
   \end{bmatrix}.
\end{equation}
Also, the predicted state variable about battery capacity mentioned in constraint (\ref{eq_Fel_constrain}) is derived as 
\begin{equation} \label{eq_lagu_battery}
    \begin{aligned}
     & E_{\text{s}}^{\text{min}} \leq E_{\text{s}}(t+m) \leq E_{\text{s}}^{\text{max}} \\
     & \Rightarrow E_{\text{s}}^{\text{min}} \leq \boldsymbol{D}_{s}\boldsymbol{ x}(t+m|t)\leq E_{\text{s}}^{\text{max}}\\
     & \Rightarrow E_{\text{s}}^{\text{min}} \leq \boldsymbol{D}_{s}\boldsymbol{A}^{m}\boldsymbol{ x}(t)+\boldsymbol{D}_{s}\bm{\varphi}(m)\boldsymbol{\eta}\leq E_{\text{s}}^{\text{max}}\\
     & \Rightarrow \boldsymbol{D}_{s}\bm{\varphi}(m)\boldsymbol{\eta} \leq  E_{\text{s}}^{\text{max}}-\boldsymbol{D}_{s}\boldsymbol{A}^{m}\boldsymbol{ x}(t) \\
    & -\boldsymbol{D}_{s}\bm{\varphi}(m)\boldsymbol{\eta} \leq
    \boldsymbol{D}_{s}\boldsymbol{A}^{m}\boldsymbol{ x}(t)-E_{\text{s}}^{\text{min}},
    \end{aligned}
\end{equation}
where $\boldsymbol{D}_{s}=\begin{bmatrix}1 & 0\end{bmatrix}$.

Constraints  (\ref{eq_lagu_umax})--(\ref{eq_lagu_battery}) are combined into a more concise form: 
\begin{equation}\label{eq_Acstn_m}
\boldsymbol{A}_{\text{cstn}}(m)\boldsymbol{\eta} \leq \boldsymbol{b}_{\text{cstn}}(m) \quad
m=0, 1, \ldots, M-1,
\end{equation}
where
\begin{equation*}
\boldsymbol{A}_{\text{cstn}}(m) =
\begin{bmatrix}
\boldsymbol{ L}(m)^{T} & \boldsymbol{o}_{1 \times J} &\cdots& \boldsymbol{o}_{1 \times J}\\
\boldsymbol{o}_{1 \times J} & \boldsymbol{ L}(m)^{T} &\cdots&  \boldsymbol{o}_{1 \times J} \\
\vdots &\vdots & \ddots &\vdots\\
\boldsymbol{ o}_{1 \times J} & \boldsymbol{ o}_{1 \times J} &\cdots &\boldsymbol{ L}(m)^{T} \\
-\boldsymbol{ L}(m)^{T} & \boldsymbol{ o}_{1 \times J} &\cdots& \boldsymbol{ o}_{1 \times J}\\
 \boldsymbol{o}_{1 \times J} & -\boldsymbol{ L}(m)^{T} &\cdots&  \boldsymbol{ o}_{1 \times J} \\
\vdots &\vdots & \ddots &\vdots\\
 \boldsymbol{o}_{1 \times J} & \boldsymbol{ o}_{1 \times J} & \cdots & -\boldsymbol{ L}(m)^{T} \\
\multicolumn{4}{c}{\boldsymbol{D}_{s}\bm{\varphi}(m)}\\
\multicolumn{4}{c}{-\boldsymbol{D}_{s}\bm{\varphi}(m)}
\end{bmatrix},
\end{equation*}
and
\begin{equation*}
\boldsymbol{b}_{\text{cstn}}(m)=
\begin{bmatrix}
S^{\text{max}}\\
\gamma_{c_{1}}^{\text{max}}\\
\vdots\\
\gamma_{c_{\ell}}^{\text{max}}\\
S^{\text{max}}\\
-\gamma_{c_{1}}^{\text{min}}\\
\vdots\\
-\gamma_{c_{\ell}}^{\text{min}}\\
E_{\text{s}}^{\text{max}}-\boldsymbol{D}_{s}\boldsymbol{A}^{m}\boldsymbol{ x}(t)\\
\boldsymbol{D}_{s}\boldsymbol{A}^{m}\boldsymbol{ x}(t)-E_{\text{s}}^{\text{min}}
\end{bmatrix}.
\end{equation*}
These  inequalities in (\ref{eq_Acstn_m}) represent the feasible region to the optimization problem. To efficiently search for solutions within this feasible region, a convex sampler technique is employed.

Finally, all constraints are combined into a single canonical form as
\begin{equation}\label{eq_Acstn_plum}
\boldsymbol{A}_{\text{cstn}}^{'}\boldsymbol{\eta} \leq \boldsymbol{b}_{\text{cstn}}^{'},
\end{equation}
where
\begin{equation*}
\boldsymbol{A}_{\text{cstn}}^{'}=
\begin{bmatrix}
\boldsymbol{A}_{\text{cstn}}(0)\\
\boldsymbol{A}_{\text{cstn}}(1)\\
\vdots\\
\boldsymbol{A}_{\text{cstn}}(M-1)
\end{bmatrix},\text{ and }\boldsymbol{b}_{\text{cstn}}^{'}=\begin{bmatrix}
\boldsymbol{b}_{\text{cstn}}(0)\\
\boldsymbol{b}_{\text{cstn}}(1)\\
\vdots\\
\boldsymbol{b}_{\text{cstn}}(M-1)
\end{bmatrix}.
\end{equation*}

With all the mathematical transformations and reduction in the number of decision variables using the Laguerre function, the final multiobjective optimization problem is represented as follows: 
\begin{equation}\label{eq_multiobjective_eta}
    \begin{split}
     &\mathop{\min}_{\substack{
     \boldsymbol{\eta}, t_{b}\\
     b\in \mathcal{A_{\text{tf}}}
     }}\text{ }F_{\text{cost}}\\
     &\mathop{\min}_{\substack{
     \boldsymbol{\eta}, t_{b}\\
     b\in \mathcal{A_{\text{tf}}}
     }}\text{ }F_{\text{dissat}}\\
     &\text{subject to }~(\ref{eq_Xb_t})\text{ and } ~(\ref{eq_Acstn_plum}).
    \end{split}
\end{equation}
In this model, the number of decision variables is $\left|\mathcal{A_{\text{tf}}} \right|+(\left|\mathcal{A_{\text{pf}}} \right|+1)J$. While the number of decision variables is compatible with that in the original multiobjective optimization problem (\ref{eq_multiobjective}),
 this new formulation in (\ref{eq_multiobjective_eta}) allows for 
a more efficient exploration of the feasible set. Originally, decision variables involve some dynamical constraints, such as the energy storage dynamics in (\ref{eq_Fel_tadd1}),
which are difficulty to address through 
 random generation of points by conventional MOEAs.
 By contrast, this new formulation involves decision variables in an inequality constraint (\ref{eq_Acstn_plum}), where feasible points can be  randomly generated using a convex sampler technique.

 In~(\ref{eq_multiobjective_eta}), the power control variables associated with the power-flexible appliances and energy storage system 
 have been replaced by Laguerre coefficients.
 $L\leq M$ can be set to reduce the number of 
 decision variables. However, if $L$ is too small,
the accuracy of approximating the control signals $P_s(t)$ and   $P_c(t)$
can be degraded. The best practice is to set 
$L$ slightly lower than $M$ for a small reduction while the new problem formulation can 
facilitate efficient exploration of the feasible set. 

By solving (\ref{eq_multiobjective_eta}), the proposed MPC-based method effectively manages model mismatch by integrating the receding horizon principle with real-time feedback correction and explicit constraint handling. Specifically, at each time step, newly observed system states—such as battery energy level, electricity price, renewable generation, and power load—are incorporated into the optimization problem, ensuring that the predicted trajectories accurately reflect the latest system conditions. 
This is accomplished by updating the system state 
$\boldsymbol{x}(t)$  using real-time measurements, refining the latest estimates of renewable generation, electricity prices, and power load, and integrating these updates into the optimization process to correct deviations from prior predictions.

The feedback mechanism plays a crucial role in mitigating model mismatch by dynamically adjusting control actions based on the discrepancy between predicted and actual system behavior. This is done by solving the optimization problem in a receding horizon fashion, where each new iteration leverages updated state estimates to refine future predictions. Furthermore, model errors are implicitly corrected through the inclusion of state feedback in the cost function and constraints, allowing the controller to adapt its decisions in real time. By continuously updating the state-space representation and adjusting control inputs in response to observed deviations, the proposed method enhances robustness against uncertainties, ensuring optimal performance under varying conditions.

\subsection{Algorithm Design and Proposed Multiobjective  Model Predictive Control}\label{subsec_ProposedMOEA}
An MOEA is developed to solve the multiobjective optimization problem. Conventional algorithms use penalty methods to address physical constraints; 
infeasible points are gradually removed from the solution process, which can be challenging 
if the number of decision variables is large 
and random samples mostly produce infeasible points.  
Hence, in this study, the convex sampler method is adopted to generate feasible solutions. Based on this method, the  proposed algorithm can search for solutions that satisfy the linear inequality constraints in (\ref{eq_Acstn_plum}). 

At first, the population at iteration $It$ is defined as
\begin{equation*}
  \mathcal{H}_{It}=\{\boldsymbol{h}_{It,1}, \boldsymbol{h}_{It,2}, \ldots, \boldsymbol{h}_{It,N_{\text{pop}}} \}
\end{equation*}
where the chromosome $\boldsymbol{h}_{It,n}$ is divided into two parts
\begin{equation*}
\boldsymbol{h}_{It,n}=
    \begin{bmatrix}
     t_{b,n}& \boldsymbol{\eta}_{n}
    \end{bmatrix} \quad n=1, 2, \ldots, N_{\text{pop}}.
\end{equation*}
The first part consists of discrete variables related to start times for the time-flexible appliances.
The second part consists of continuous variables considering the battery dynamic and power-flexible appliances.

Algorithm~\ref{alg_Solution} is proposed to generate initial solutions.
The proposed MOEA algorithm adopts the convex sampler method in Algorithm~\ref{alg_Convex} and the improved algorithm based on convex sampler in Algorithm~\ref{alg_ImprovedConvex} to search in the feasible set.
The crossover and mutation operation in the proposed MOEA algorithm are introduced in Algorithm~\ref{alg_Crossover} and Algorithm~\ref{alg_Mutation}.
Algorithm~\ref{alg_Proposed} is the pseudocode of the proposed MOEA algorithm. This algorithm can solve constrained multiobjective problem and reduce uncertainty influence from the forecasting error. Further details are presented in the following sections.

In Algorithm~\ref{alg_Solution}, 
population $\mathcal{H}_{0}$ is the set of better chromosomes during the evolutionary process;
population $\bar{\mathcal{H}}$ is responsible for supporting the crossover operation to be applied at a later stage.
In Step 1),
the initial population $\mathcal{H}_{0}$ is produced after iterations. In Step 1.1), the start time $t_{b,n}$ for the time-flexible appliances must be decided first ensuring Constraint~(\ref{eq_Xb_t}) is satisfied. In Step 1.2), the linear inequalities, $\boldsymbol{A}_{\text{cstn}}^{'}$ and $\boldsymbol{b}_{\text{cstn}}^{'}$ can be defined~(\ref{eq_Acstn_plum}) after the power consumption of time-flexible appliances is known. The set of feasible solutions satisfy the following equation after determining the power consumption of time-flexible appliances
\begin{equation}\label{eq_S}
\mathcal{S}=
\{ \boldsymbol{\eta} | \boldsymbol{A}_{\text{cstn}}^{'} \boldsymbol{\eta}\leq \boldsymbol{b}_{\text{cstn}}^{'} \}.
\end{equation}
In Step 1.3),
the initial feasible point $\boldsymbol{\eta}_{ini}$ 
belonging to  ~(\ref{eq_S}) 
is randomly generated. 
In Step 1.4), $\boldsymbol{\eta}_{n}$ is randomly generated by Algorithm~\ref{alg_Convex}.
In Step 1.5),  feasible solutions $\boldsymbol{h}_{0,n}$ are obtained by combining $t_{b,n}$ and $\boldsymbol{\eta}_{n}$.
The population $\mathcal{H}_{0}$ is now initialized and 
population $\bar{\mathcal{H}}$ is produced following Steps 2) and 3).
In Step 2), the chromosomes are produced by random generation in~(\ref{eq_lagu_combine}).  Step 3) calculates the distances between chromosomes 
and remove those close to each other, e.g., using 
the crowding distance sorting method\cite{10Pindoriya}, to keep a manageable solution set while ensuring the solution diversity.
\begin{algorithm}
\caption{Solution Initialization}
\label{alg_Solution}
\begin{algorithmic}
\Require  Parameters of appliances $\alpha_{a}$, $\beta_{a}$, $\gamma_{a}$, $\alpha_{b}$, $\beta_{b}$, $T_{b}$, $\gamma_{b}$, $\alpha_{c}$, $\beta_{c}$, $\gamma_{c}^{\text{min}}$, $\gamma_{c}^{\text{max}}$, the battery parameter $\rho$, $S^{\text{max}}$, $E_{\text{s}}^{\text{max}}$, predicted information $P_{\text{re}}(t+m)$, $P_{a}^{\text{con}}(t+m)$ and population size $N_{\text{pop}}$. 
\Ensure  $\mathcal{H}_{0}$ and $\bar{\mathcal{H}}$
\State  \textbf{Step 1)} Initialize the population $\mathcal{H}_{0}$ by complex sampler method.
\State  \textbf{For} $n=1:N_{\text{pop}}$
        \State \quad \textbf{Step 1.1)} Initialize the start time of all time-flexible
        \State \quad appliances $t_{b}$ in ~(\ref{eq_Xb_t}). 
        \State \quad \textbf{Step 1.2)} Produce the linear inequalities $\boldsymbol{A}_{\text{cstn}}^{'}$ and $\boldsymbol{b}_{\text{cstn}}^{'}$.
        \State \quad \textbf{Step 1.3)} Generate the first initial feasible solution $\boldsymbol{\eta}_{ini}$.
        \State \quad \textbf{Step 1.4)} Initialize $\boldsymbol{\eta}_{n}$ by Algorithm~\ref{alg_Convex} with given position \State \quad $\mathcal{E}_{1}$ and initial feasible solution $\boldsymbol{\eta}_{ini}$.
        \State \quad \textbf{Step 1.5)} Initialize feasible solutions $\boldsymbol{h}_{0,n}=\begin{bmatrix}t_{b,n} & \boldsymbol{\eta}_{n}\end{bmatrix}$
\State  \textbf{End for}
\State  \textbf{Step 2)}  Generate  chromosomes randomly. 
\State  \textbf{Step 3)} 
Keep $\bar{\mathcal{H}}$ at a manageable size of   $N_{\text{pop}}$ by removing
some chromosomes if necessary.
\end{algorithmic}
\end{algorithm}

The convex sampling methods in Algorithms~\ref{alg_Convex} and ~\ref{alg_ImprovedConvex} are applied to generate feasible solutions.
In  Steps $1)$ and $2)$ of Algorithm~\ref{alg_Convex}, the distance of the direction vector $i$ is calculate to determine if the direction is positive or negative.
In Step $2.1)$, 
 $d_{\text{pos}}$ and $d_{\text{neg}}$ are 
 updated after calculating all the distances to the constraints. 
In Step $3)$, the new feasible vector $\boldsymbol{\eta}_{n}$ 
is obtained by changing some of its elements, which is controlled by the input set  $\mathcal{E}$.
A similar procedure is performed in  
Algorithm~\ref{alg_ImprovedConvex}
except that 
a direction vector $\boldsymbol{\eta}_{dir}$ is 
used to control 
the search direction.

The purpose of introducing two versions of convex sampling is to diversify the solutions in the mutation operation presented in Algorithm~\ref{alg_Mutation}.
The mutation introduces random changes to individual solutions (chromosomes). By altering one or more genes (variables) in a solution, mutation helps explore new areas of the search space that 
have not been previously visited.
 Mutation adds diversity to the population and ensures that the algorithm does not converge prematurely to a suboptimal solution.

\begin{algorithm}
\caption{Convex sampler I}
\label{alg_Convex}
\begin{algorithmic}
\Require  $\boldsymbol{A_{\text{cstn}}}^{'}$, $\boldsymbol{b_{\text{cstn}}}^{'}$, $\mathcal{E}$, $It$ and $\boldsymbol{\eta}_{n}$
\Ensure $\boldsymbol{\eta}_{n}^{'}$
\State \textbf{For} $i\in \mathcal{E}$
    \State \quad \textbf{Step 1)} Calculate the number of  constraints $N_{\text{c}}$; Let
    \State \quad $n_{\text{c}}=1$.
    \State \quad \textbf{Step 2)} Calculate the distance to the $n_{\text{c}}^{th}$ constraint to
    \State \quad determine positive and negative direction.
    \State \quad \textbf{while} $n_{\text{c}} \leq N_{\text{c}}$ \textbf{do}
        \State \quad\quad \textbf{Step 2.1)} Update the  largest distances associated with \State \quad\quad positive direction $d_{\text{pos}}$ and  negative direction $d_{\text{neg}}$;
        \State \quad\quad $n_{\text{c}}:=n_{\text{c}}+1$.
    \State \quad \textbf{end while}
    \State \quad \textbf{Step 3)} Update   $\boldsymbol{\eta}_{n}$ by
    \State \quad \begin{equation*}
                   \begin{bmatrix}\boldsymbol{\eta}_{n}\end{bmatrix}_{i}=\begin{bmatrix}\boldsymbol{\eta}_{n}\end{bmatrix}_{i}-d_{\text{neg}}+(d_{\text{pos}}+d_{\text{neg}})C_{\text{rand}}
                 \end{equation*}
    \State \quad where $\begin{bmatrix}\boldsymbol{\eta}_{n}\end{bmatrix}_{i}$ represents the $i$th element of $\boldsymbol{\eta}_{n}$
    \State \quad and random number $C_{\text{rand}}$ is defined as
    \State \quad \text{ }\begin{equation*}
                    C_{\text{rand}}\!:=\!
                    \begin{cases}
                        \text{randomly from }\{0, 1\},\!&\text{if }It\equiv 0\!\pmod 5;\\
                        \text{randomly from }[0, 1],\!&\text{otherwise}.
                    \end{cases}
                  \end{equation*}
\State \textbf{End for}
\end{algorithmic}
\end{algorithm}
\begin{algorithm}
\caption{Convex Sampler II}
\label{alg_ImprovedConvex}
\begin{algorithmic}
\Require  $\boldsymbol{A_{\text{cstn}}}^{'}$, $\boldsymbol{b_{\text{cstn}}}^{'}$ and $\boldsymbol{\eta}_{n}$
\Ensure $\boldsymbol{\eta}_{n}^{'}$
\State \textbf{Step 1)} Calculate the direction vector
\begin{equation*}
\boldsymbol{\eta}_{dir}=
          \boldsymbol{\eta}_{\text{rand}}-\boldsymbol{\eta}_{n}
\end{equation*}
where $\boldsymbol{\eta}_{\text{rand}}$ is a random point.
\State \textbf{Step 2)} Calculate the number of the constraints $N_{\text{c}}$; Let $n_{\text{c}}=1$.
\State \textbf{Step 3)} Calculate the distance to the $n_{\text{c}}^{th}$ constraint for positive and negative direction.
\State \textbf{while} $n_{\text{c}} \leq N_{\text{c}}$ \textbf{do}
        \State \quad \textbf{Step 3.1)} 
        Update the  largest distances associated with \State \quad positive direction $d_{\text{pos}}$ and  negative direction $d_{\text{neg}}$;
        \State \quad
        $n_{\text{c}}:=n_{\text{c}}+1$.
\State \textbf{end while}
\State \textbf{Step 4)} Update the solution
\begin{equation*}
\boldsymbol{\eta}_{n}^{'}=\boldsymbol{\eta}_{n}+(-d_{\text{neg}}+(d_{\text{pos}}+d_{\text{neg}})C_{\text{rand}})\boldsymbol{\eta}_{dir}
\end{equation*}
where $C_{\text{rand}}\in [0,1]$ is a random number.
\end{algorithmic}
\end{algorithm}

\begin{algorithm}
\caption{Mutation }
\label{alg_Mutation}
\begin{algorithmic}
\Require $\mathcal{H}_{It}$ and $It$
\Ensure $\boldsymbol{h}_{It}^{'}$
\State \textbf{Step 1)} Randomly select one of the solution $\boldsymbol{h}_{It,n}$ in the population $\mathcal{H}_{It}$.
\State \textbf{Step 2)} Keep the start time $t_{b}$. Randomly apply Algorithm~\ref{alg_ImprovedConvex}
or Algorithm~\ref{alg_Convex} to generate 
$\boldsymbol{\eta}_{n}$.
\State \textbf{Step 3)}
From 
$\boldsymbol{h}_{It}^{'}$ using 
$t_{b}$ and $\boldsymbol{\eta}_{n}$.
\end{algorithmic}
\end{algorithm}

\begin{algorithm}
\caption{Crossover}
\label{alg_Crossover}
\begin{algorithmic}
\Require $\mathcal{H}_{It}$ and $\bar{\mathcal{H}}$
\Ensure  $\boldsymbol{h}_{It}^{'}$ and $\boldsymbol{h}_{It}^{''}$
\State \textbf{Step 1)} Randomly select one of the chromosomes $\boldsymbol{h}_{It,n}$ in the population $\mathcal{H}_{It}$.
\State \textbf{Step 2)} Randomly select one of the chromosomes $\bar{\boldsymbol{h}}$ in the population $ \mathcal{H}_{It}\cup \bar{\mathcal{H}}$.
\State \textbf{Step 3)} Randomly generate new starting time $t_{b}^{'}$ and $t_{b}^{''}$ .
\State \textbf{Step 4)} Assign
\begin{equation*}
    \begin{split}
  &\boldsymbol{h}_{It}^{'}=\begin{bmatrix}
     t_{b}^{'} & C_{\text{rand}} \boldsymbol{\eta}_{n}+ (1-C_{\text{rand}})\bar{\boldsymbol{\eta}}
    \end{bmatrix}\\
  &\boldsymbol{h}_{It}^{''}=\begin{bmatrix}
     t_{b}^{''} & (1-C_{\text{rand}}) \boldsymbol{\eta}_{n}+ C_{\text{rand}}\bar{\boldsymbol{\eta}}
    \end{bmatrix}
    \end{split}
\end{equation*}
where $C_{\text{rand}}\in [0,1]$ is a uniform random number. 
\end{algorithmic}
\end{algorithm}

To
refine and improve existing solutions by searching in the vicinity of known good solutions,
Algorithm~\ref{alg_Crossover}
presents 
the crossover operator that combines parts of two parent solutions to produce offspring solutions. Crossover is designed to exploit the information already present in the population by mixing and matching high-quality traits from the parents. This operator helps the algorithm exploit the most promising areas of the search space by generating new solutions that inherit desirable characteristics from the parents.

Algorithm~\ref{alg_Proposed} presents the proposed MOMPC
for residential demand response management, utilizing the aforementioned algorithms. 
The algorithm iteratively 
searches the feasible space through exploration and exploitation realized by 
mutation in Algorithm~\ref{alg_Mutation} and crossover in Algorithm~\ref{alg_Crossover}, respectively.

\begin{algorithm}
\caption{Proposed MOMPC}
\label{alg_Proposed}
\begin{algorithmic}
\Require Parameters of appliances $\alpha_{a}$, $\beta_{a}$, $\gamma_{a}$, $\alpha_{b}$, $\beta_{b}$, $T_{b}$, $\gamma_{b}$, $\alpha_{c}$, $\beta_{c}$, $\gamma_{c}^{\text{min}}$, $\gamma_{c}^{\text{max}}$, battery parameter $\rho$, $S^{\text{max}}$, $E_{\text{s}}^{\text{max}}$, parameter of  dissatisfaction $\theta_{b}$, $t_{b}^{\text{req}}$, $k_{b}$, $\xi_{c}$, $P_{c}^{\text{nor}}$.
\Ensure $\boldsymbol{u}(t)$
    \State \textbf{Step 1)} Set the MOEA parameters $R_{\text{mu}}$, $R_{\text{co}}$, $N_{\text{pop}}$, $It_{\text{max}}$, the Laguerre parameters $p$ and $J$.
    \State \textbf{Step 2)} Update the predicted information $P_{\text{re}}(t+m)$, $P_{a}^{\text{con}}(t+m)$ and $\lambda(t+m)$.
    \State \textbf{Step 3)} Produce chromosomes $\mathcal{H}_{0}$ and $\bar{\mathcal{H}}$ in Algorithm~\ref{alg_Solution}; let $It=0$.
    \State \textbf{While} $It \leq It_{\text{max}}$ \textbf{do}
            \State \quad \textbf{Step 3.1)} Apply crossover operation in  Algorithm~\ref{alg_Crossover}. 
            \State \quad \textbf{Step 3.2)} Apply mutation operation in Algorithm~\ref{alg_Mutation}.
            \State \quad \textbf{Step 3.3)} Check solutions for feasibility and eliminate 
            \State \quad infeasible solutions.
            \State \quad \textbf{Step 3.4)} Sort rank using non-dominated sorting and
            \State \quad calculate crowding distance.
            \State \quad \textbf{Step 3.5)} Remove dominated chromosomes 
    \State \textbf{End while}
    \State  \textbf{Step 4)} Output Pareto set.
    \State  \textbf{Step 5)} Select the final solution and determine the start time if $t=\alpha_{\text{b}}$.
    \State  \textbf{Step 6)} Recover Laguerre functions and implement control signal $\boldsymbol{u}(t)$
using~(\ref{eq_ut})

\end{algorithmic}
\end{algorithm}

\section{Numerical Analysis}\label{sec_Simulation}
In this section, a home energy management system is presented. This smart home consists of $3$ inflexible appliances, $1$ shiftable home appliance and $2$ power flexible appliances. Energy storage is considered within this system and the initial energy in the battery is $4\text{kWh}$. Table~\ref{tab_parameter} shows all parameters used in the following analysis. The starting time in the day is setting at the $8$ A.M., thus the starting time and ending time $\alpha_{a_{3}}^{1\text{th}}=\beta_{a_{3}}^{1\text{th}}=1$, means these appliances are implemented in the interval $8$ A.M.-$9$ A.M. (the first time slot). 
In this paper,  energy scheduling is introduced and a day is divided into $24$ time intervals, i.e., $\triangle t=1$ and  $|\mathcal{T}|=24$. 

\begin{table}[htbp]
\caption{Parameters about home equipment}
\setlength{\tabcolsep}{1.5mm}
\renewcommand{\arraystretch}{1.3}
\centering
\begin{tabular}{|c||c|c|}
\hline
\textbf{Inflexible appliances} & \textbf{Power rating} & \textbf{Parameters}  \\
\hline
$P_{a_{1}}$ & $ 0.2 \text{kW}$ & $\alpha_{a_{1}}=1$, $\beta_{a_{1}}=24$\\
\hline
$P_{a_{2}}$ & $ 1.2 \text{kW}$ & $\alpha_{a_{2}}=24$, $\beta_{a_{2}}=24$\\
\hline
$P_{a_{3}}$ & $ 1.2 \text{kW}$ & $\alpha_{a_{3}}^{1\text{th}}=1$, $\beta_{a_{3}}^{1\text{th}}=1$,\\
 & &$\alpha_{a_{3}}^{2\text{th}}=10$, $\beta_{a_{3}}^{2\text{th}}=10$,\\
 & &$\alpha_{a_{3}}^{3\text{th}}=13$, $\beta_{a_{3}}^{3\text{th}}=13$ \\
\hline
\hline
\textbf{Shiftable appliances} & \textbf{Power rating} & \textbf{Parameters}  \\
\hline
$P_{b_{1}}$ &  0.7 \text{kW} & $\alpha_{b_{1}}=11$, $\beta_{b_{1}}=23$,\\
 & & $\theta_{b_{1}}=0.001$, $t_{b_{1}}^{\text{req}}=11$\\
 & & $T_{b_{1}}=2$, $k_{b_{1}}=3$\\
\hline
\hline
\textbf{Power flexible } & \textbf{Power rating} & \textbf{Parameters}  \\
\textbf{appliances}&&\\
\hline
$P_{c_{1}}$ & $\gamma_{c_{1}}^{\text{min}}=0.2\text{kW}$, & $\alpha_{c_{1}}=11$, $\beta_{c_{1}}=16$ \\
 & $\gamma_{c_{1}}^{\text{max}}=0.8 \text{kW}$ & $\xi_{c_{1}}=1$, $P^{\text{nor}}_{c_{1}}=0.8$\\
\hline
$P_{c_{2}}$ &  $\gamma_{c_{2}}^{\text{min}}=0 \text{kW}$, & $\alpha_{c_{2}}=1$, $\beta_{c_{2}}=24$ \\
 & $\gamma_{c_{2}}^{\text{max}}=1.4 \text{kW}$ & $\xi_{c_{2}}=0.4$, $P^{\text{nor}}_{c_{2}}=1.4$\\
\hline
\hline
\textbf{Battery} & \multicolumn{2}{c|}{\textbf{Parameters}}  \\
\hline
\hline
$\rho$ & \multicolumn{2}{c|}{$\sqrt[24]{0.9}$}  \\
\hline
$S^{\text{max}}$ & \multicolumn{2}{c|}{$3 \text{kW}$} \\
\hline
$E_{\text{s}}^{\text{max}}$ & \multicolumn{2}{c|}{$10 \text{kWh}$} \\
\hline
$E_{\text{s}}^{\text{min}}$ & \multicolumn{2}{c|}{$3 \text{kWh}$} \\
\hline
\end{tabular}
\label{tab_parameter}
\end{table}

\begin{figure}
\hspace{-0.5cm}
  \includegraphics[width=15cm,height=12cm]{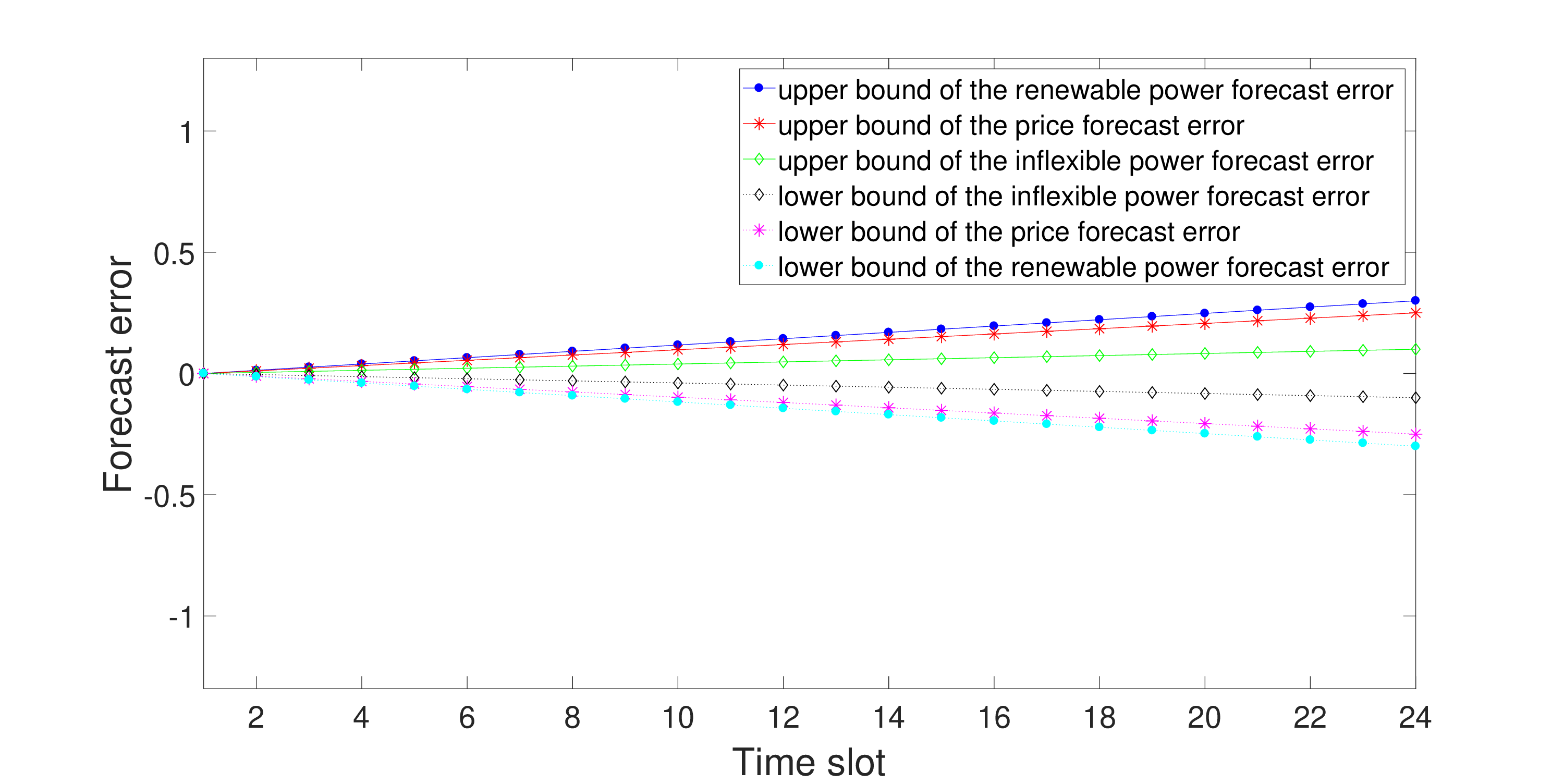}
  \caption{The bound of three different forecasting error in the current time slot.}\label{fig_forecast}
\end{figure}

\begin{figure}
\centering
\begin{equation*}
\begin{array}{c}
\includegraphics[width=15cm,height=5.2cm]{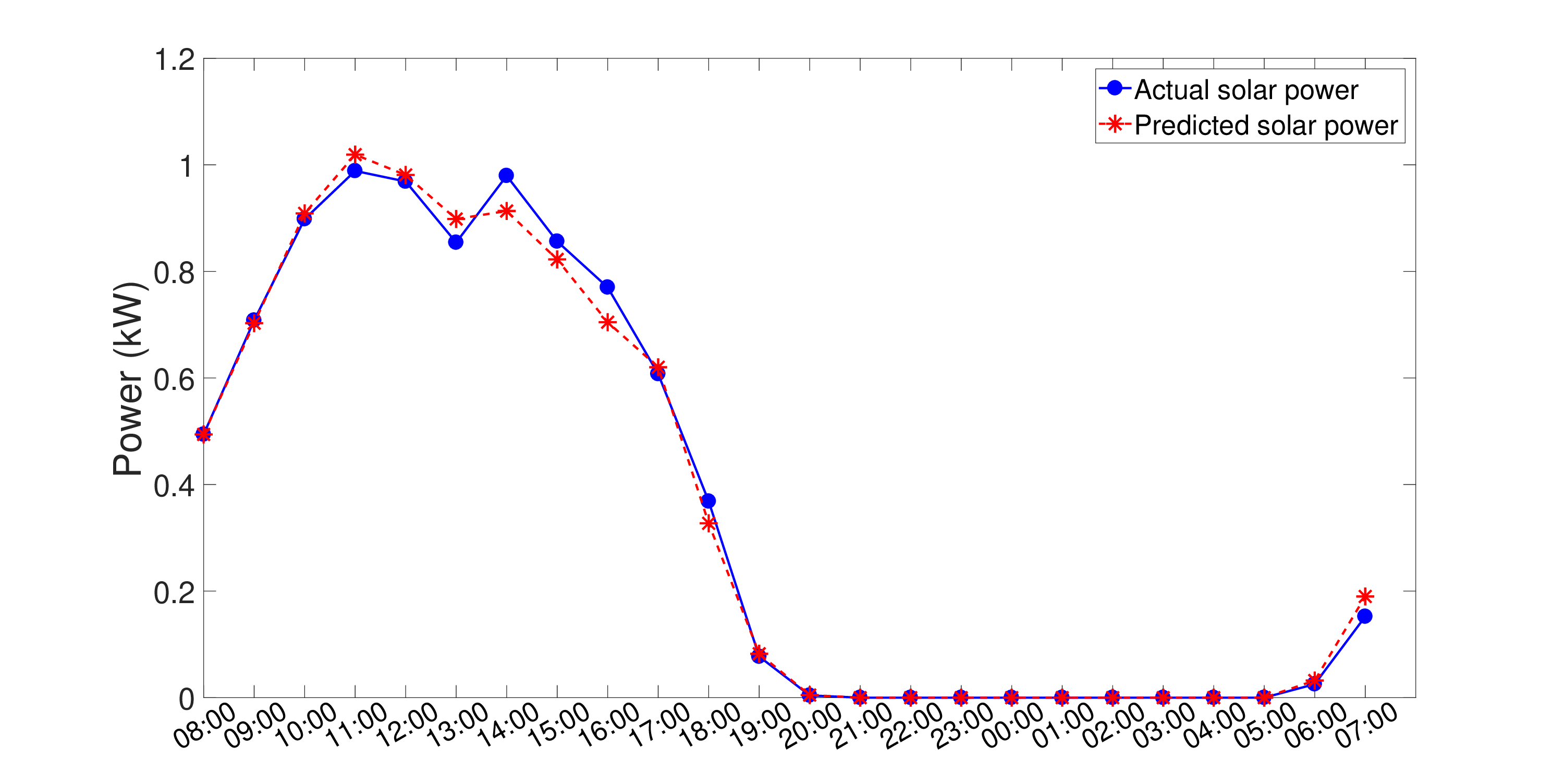} \\
\mbox{(a)} \\
\includegraphics[width=15cm,height=5.2cm]{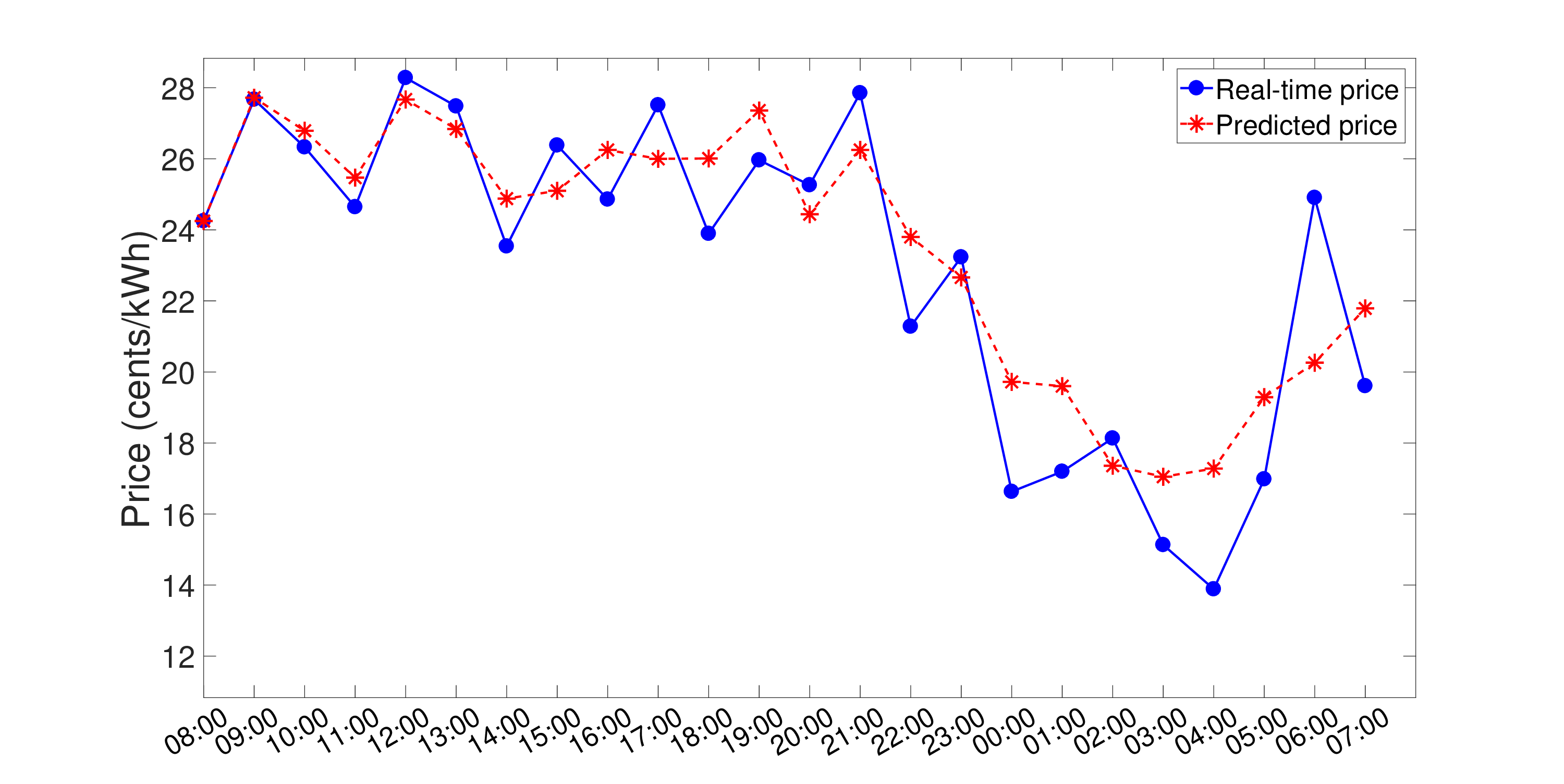} \\
\mbox{(b)} \\
\includegraphics[width=15cm,height=5.2cm]{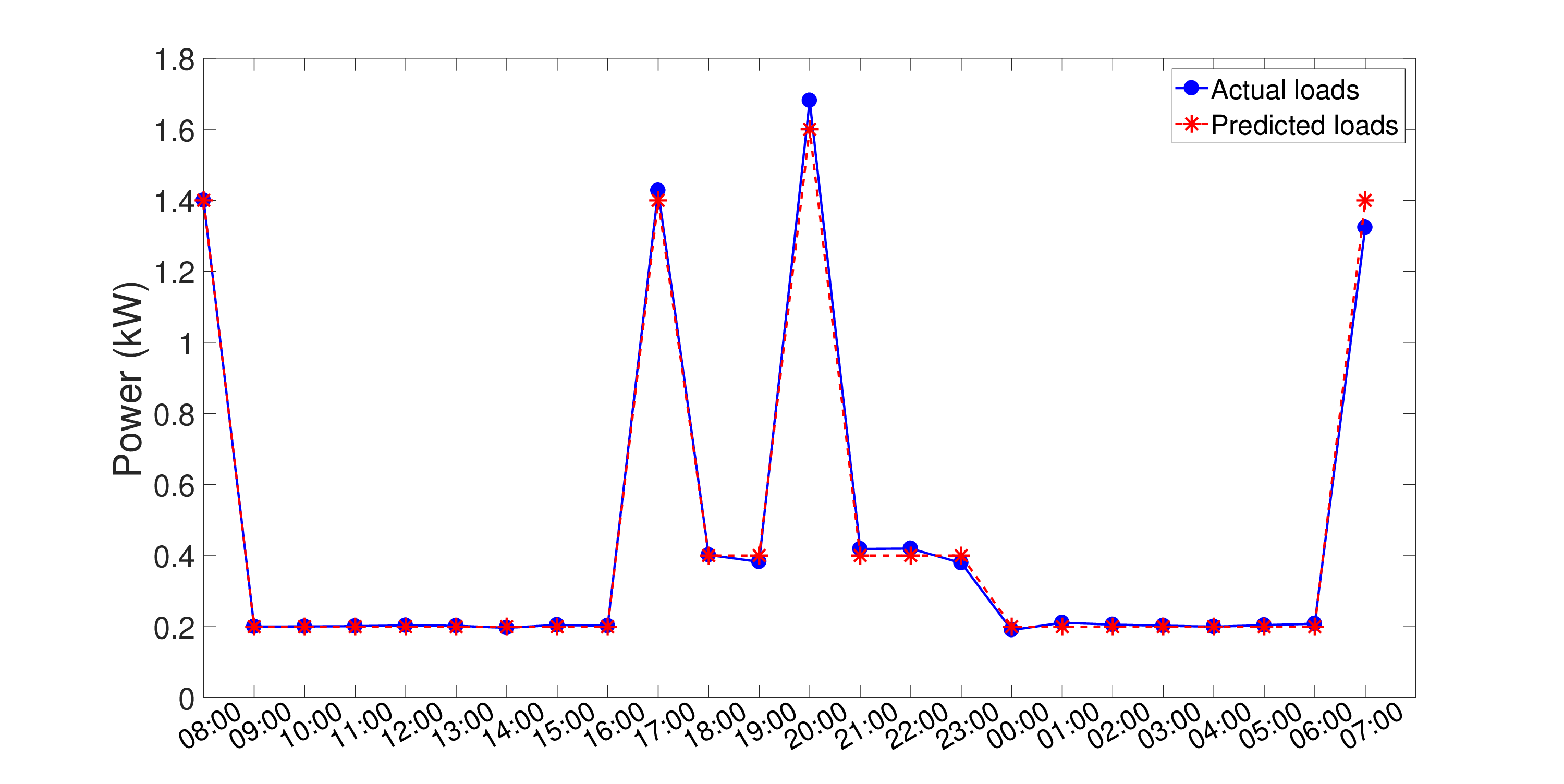} \\
\mbox{(c)}
\end{array}
\end{equation*}
\caption{Predicted and true values for (a) renewable power, (b) electricity price,
and (c) power loads.} \label{fig_uncertainty}
\end{figure}

Real-world data of solar energy and real-time pricing from PJM were used. 
Figure~\ref{fig_forecast} shows a model for the ranges of forecast errors \cite{16Zhang}.
Prediction errors occurred in renewable power forecasts, electricity price forecasts, and power load forecasts. 
The absolute values of 
these errors tend to increase as the forecast time slot moves further into the future. This is expected, as predicting values becomes increasingly challenging the further they are from the current time.
 To be specific, 
if $t<t'$, then the
absolute values of prediction errors at time slot $t'$ are larger than those at time $t$. 
In the simulations conducted, no specific methods for value prediction were employed; predicted values were generated based on the true values corrupted with random noise following the model in \cite{16Zhang}.
In practice, any state-of-the-art prediction methods,
such as 
 long short-term memory networks and temporal convolutional networks \cite{Siami2019}, can be used for  predicting $M$ steps ahead of renewable power generation and electricity prices.
Figure~\ref{fig_uncertainty} presents  samples of predicted and true data values of the renewable power, electricity price and power load.


The proposed method 
with Laguerre parameters $J=15$, $p=0.8$ and $M=20$
was compared with existing multiobjective methods for demand response management: NSGA-II with the penalty method \cite{09Pedrasa}, constrained NSGA-II \cite{12Salinas}, PODR \cite{20Chiu}, and MOIA \cite{15Chiu}
for which the common parameters 
$N_{\text{pop}}=200$, $It_{\text{max}}=1000$, $R_{\text{co}}=0.2$, and $R_{\text{mu}}=0.8$
were used. 
Comparable methods were adopted while preserving their original methodologies as much as possible, with only slight modifications to suit our simulation scenarios.
After Pareto fronts were obtained, knee solutions were selected for comparison \cite{15Chiu}.

The hardware used in this study includes an AMD 7955WX CPU (16 cores, 32 threads, 4.5 GHz), 128 GB of ECC RAM, an NVIDIA T400 GPU with 4 GB of GDDR6 memory, and a 1 TB PCIe-4x4 2280 TLC SSD for primary storage. The computational time required to generate control signals for each time slot ranged from 16.40 to 17.72 minutes, with an average of 16.40 minutes and a standard deviation of 0.62 minutes. Notably, this computational time could be reduced further with more advanced hardware.

 Although this study assumes an hourly real-time pricing market, a finer time resolution, such as a 30-minute interval, has also been explored in the literature and can be effectively handled by the proposed method.
 In practice, the algorithm generates high-level control signals to govern the operation of energy storage systems and appliances. At a lower level, power electronics would implement these operational commands with finer time resolutions.

\subsection{Pareto Fronts and Convergence Analysis}

\begin{figure}
  \includegraphics[width=15cm]{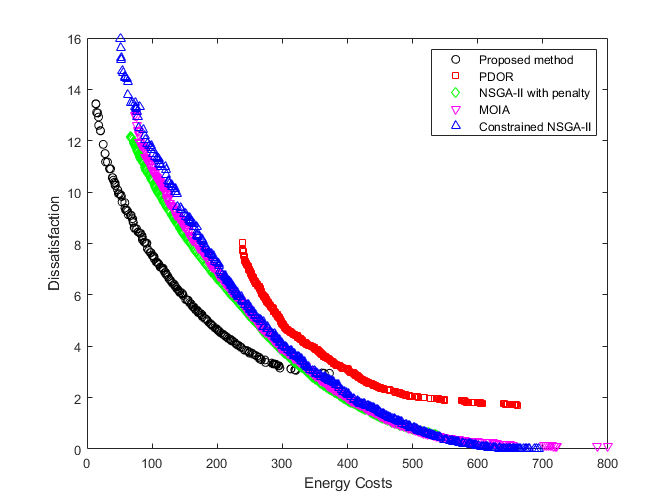}
  \caption{Pareto fronts obtained by different methods for the minimization of energy costs (x-axis) and user dissatisfaction (y-axis). The proposed method (black circles) employs a model predictive control framework to mitigate uncertainty, resulting in superior tradeoff performance compared to other methods, including PDOR (red squares), NSGA-II with penalty (green diamonds), constrained NSGA-II  (blue triangles), and MOIA (magenta inverted triangles).}\label{fig_Pareto}
\end{figure}

In multiobjective optimization, the Pareto front represents the set of non-dominated solutions that balance tradeoffs between conflicting objectives. A solution is considered Pareto-dominant if it improves at least one objective without worsening others, making Pareto fronts a fundamental tool for evaluating solution quality \cite{coello2007evolutionary}. In the context of residential demand response, both electricity cost and user dissatisfaction must be minimized, making the position and shape of the Pareto front a key indicator of an optimization method’s effectiveness.

A well-formed Pareto front that shifts toward the bottom-left in the cost-dissatisfaction space signifies superior performance, as it offers a wider range of high-quality tradeoffs between economic efficiency and user comfort. This is particularly critical in real-time pricing scenarios, where different households may prioritize cost savings or comfort differently. By comparing Pareto fronts obtained from different methods, we can assess how well an approach navigates the tradeoff space, ensuring that no objective is unnecessarily sacrificed. Thus, Pareto dominance provides an essential benchmarking tool, allowing decision makers to select strategies that best suit user preferences.

Figure~\ref{fig_Pareto} compares Pareto fronts obtained by various demand response management methods. 
It is worth mentioning that optimization problems in (19) and (33) have different
decision variables, and the number of decision variables in (33) may differ from that in (19), even though
(33) serves as an approximation of the original problem (19).
The proposed method solving 
(33)
demonstrates a well-spread Pareto front with  lower dissatisfaction levels for equivalent energy costs compared to other methods solving (19). This performance is particularly pronounced in the low-cost region, where the proposed method consistently outperforms others in balancing the tradeoff between the two objectives. In contrast, other methods, such as PDOR, NSGA-II with penalty, constrained NSGA-II, and MOIA, exhibit higher user dissatisfaction for similar energy costs or fail to achieve the same level of tradeoff optimization.

The effectiveness of the proposed method can be attributed to its adoption of a MPC framework, which explicitly addresses uncertainties in the optimization process. By incorporating forecasts of renewable energy generation, power demand, and electricity prices, the proposed method dynamically adjusts decisions over an optimization window, leading to better overall outcomes. This adaptive approach allows the proposed method to maintain lower user dissatisfaction while achieving competitive energy costs, even in the presence of uncertainty. In contrast, other methods directly solve the original  multiobjective problem in (\ref{eq_multiobjective}) but lack mechanisms to handle uncertainties dynamically, resulting in suboptimal performance when actual conditions deviate from expectations.

Methods like PDOR and NSGA-II variants exhibit limitations in managing uncertainty. PDOR, for instance, likely employs a static optimization framework, making it less effective in scenarios where variability plays a significant role. This limitation is reflected in its higher dissatisfaction levels for equivalent energy costs. Similarly, NSGA-II with penalty and constrained NSGA-II, despite being evolutionary algorithms capable of solving deterministic problems effectively, fail to address uncertainty explicitly, leading to clustered and less effective Pareto fronts. MOIA, on the other hand, appears to emphasize diversity in its solutions but lacks explicit mechanisms to incorporate uncertainty, which negatively impacts its ability to optimize the tradeoff between objectives.

\begin{figure}
\hspace{-0.5cm}
  \includegraphics[width=15cm]{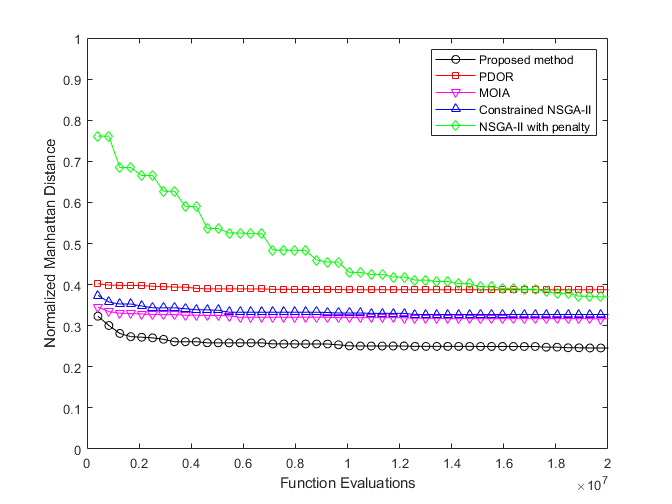}
  \caption{Normalized Manhattan distance of Pareto fronts to the ideal vector over function evaluations for various demand response management methods, illustrating the convergence behavior of each method. }\label{fig_MD}
\end{figure}

The normalized Manhattan distance  quantifies how closely a given Pareto front approaches an ideal solution, where both cost and dissatisfaction are minimized. Unlike alternative distance metrics such as the Euclidean distance, the Manhattan distance (L1 norm) is particularly well-suited for multiobjective problems because it independently measures deviations along each objective axis, preventing one objective from dominating the metric. By normalizing the distance, we ensure that differences in the scales of cost and dissatisfaction do not skew the assessment, allowing for a fair comparison between methods \cite{yu2022survey}.

 In demand response applications, where uncertainty in energy prices and user behavior can impact optimization outcomes, a lower normalized Manhattan distance indicates that a method consistently finds near-optimal tradeoffs across different scenarios. This is particularly valuable for real-world deployment, as energy management strategies must be both effective and robust to unpredictable fluctuations. By evaluating the
 normalized Manhattan distance
 across various uncertainty levels,   an algorithm’s ability to converge reliably to high-quality solutions can be assessed, ensuring practical applicability in dynamic home energy management systems.

Figure~\ref{fig_MD} 
assesses the convergence of the comparable methods by depicting the normalized Manhattan distance of the Pareto fronts from an ideal vector that represents the joint minimum energy cost and dissatisfaction.
The proposed method demonstrates the fastest and most consistent convergence, stabilizing at the lowest Manhattan distance early in the evaluation process. In contrast, PDOR converges more slowly and stabilizes at a higher Manhattan distance, reflecting its inability to achieve solutions as close to the ideal vector. MOIA  and constrained NSGA-II  exhibit moderate convergence but fail to match the performance of the proposed method, stabilizing at intermediate Manhattan distances. Notably, NSGA-II with penalty  shows the poorest performance, with the highest initial Manhattan distance and the slowest rate of improvement, stabilizing far from the ideal.

The observed deviation in Pareto fronts is consistent with the fastest and most consistent convergence
in Manhattan distances; they 
further highlight the advantages of the proposed method.
Laguerre functions are used for approximating the power control signals, which enhance the exploration of the feasible set.
 While it approximates the original  multiobjective problem, 
the method prioritizes mitigating uncertainty through the receding horizon principle, a feature absent in traditional MOEAs. 
This ensures the proposed method is more robust to real-world conditions, resulting in practical solutions with superior performance.

\subsection{Case Study and Average Performance}

\begin{figure}
\hspace{-0.5cm}
\includegraphics[width=15cm,height=12cm]{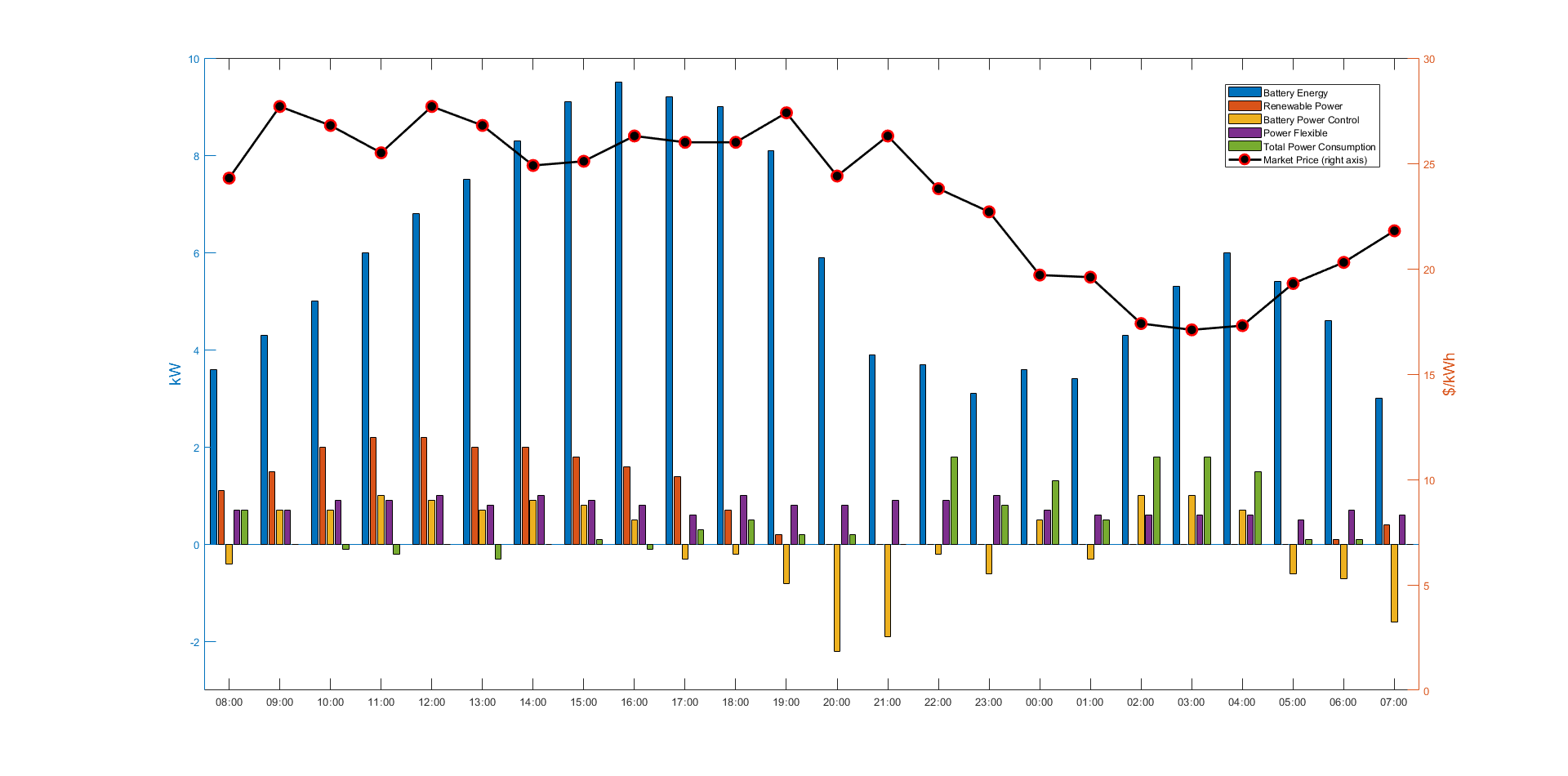}
  \caption{Power control by the proposed method for residential demand response management. The figure shows power values (left y-axis) and electricity market prices (right y-axis) over time, illustrating how the proposed method minimizes energy costs while maintaining user satisfaction. Negative total power consumption indicates excessive power sold back to the power grid with a feed-in tariff.   }\label{fig_case_study}
\end{figure}

Figure \ref{fig_case_study} illustrates the operation of a residential energy management system designed to optimize energy costs and minimize user dissatisfaction by managing a home battery storage system and power-flexible appliances. The electricity market prices, represented by the black line with red markers, exhibit significant fluctuations, peaking during certain hours (e.g., 9:00--19:00) and dipping during early morning hours (e.g., 02:00-–05:00). The proposed approach leverages these price fluctuations by strategically controlling battery energy and power consumption.

The blue bars, representing battery energy, show that the system relies heavily on battery storage to supply power during high-price periods, thereby reducing dependency on expensive grid electricity. The orange bars indicate the contribution of renewable power, which varies throughout the day due to its inherent intermittency. The yellow bars, denoting battery power control, highlight active charging and discharging patterns. During periods of high market prices, the battery discharges to offset grid consumption, while during low-price periods, it charges to store energy for future use.

The purple bars represent adjustments to the power consumption of flexible appliances. 
The normal power consumption is 2.2 kW in total. These adjustments are minor but occur strategically during high-price periods to further reduce costs without significantly affecting user satisfaction. The green bars, showing total power consumption, reveal that the system rarely sells excess power back to the grid due to the low feed-in tariff rates compared to market prices. Instead, the system prioritizes consuming stored energy or aligning consumption with low-price periods.

The proposed approach effectively reduces energy costs by aligning power usage with market price variations and optimizing battery usage. By strategically discharging the battery during high-price periods---except when it was charged earlier in the day using surplus renewable energy—--and charging it during low-price periods, the system effectively minimizes grid power purchases. Additionally, power consumption adjustments for flexible appliances are kept small and infrequent, ensuring that user dissatisfaction remains low. Overall, this approach demonstrates an efficient balance between cost savings and maintaining user satisfaction, while also incorporating renewable energy and battery storage for a more sustainable energy management solution.

\begin{table}
\caption{Average Values of Energy Costs and User Dissatisfaction}
\centering
\setlength{\tabcolsep}{1.5mm}
\renewcommand{\arraystretch}{1.5}
\begin{tabular}{|c||c|c|}
\hline
\multicolumn{3}{|c|}{\textbf{1st Scenario (without prediction errors)}}\\
\hline
\textbf{Methods} & \textbf{Energy Costs} & \textbf{Dissatisfaction} \\
\hline
 \text{Constrained NSGA-II}& 409.83 (2.69\%) & 7.48 \\
\hline
 \text{NSGA-II}& 401.51 (0.6\%) & 7.49  \\
 \text{with penalty} &  &   \\
\hline
\text{PODR} & 420.33 (5.32\%) & 7.49 \\
\hline
\text{MOIA} & 400.50 (0.35\%) & 7.47 \\
\hline
\text{Proposed method} & 399.11  & 7.49 \\
\hline
\hline
\multicolumn{3}{|c|}{\textbf{2nd Scenario (with prediction errors)}}\\
\hline
\textbf{Methods} & \textbf{Energy Costs} & \textbf{Dissatisfaction} \\
\hline
 \text{Constrained NSGA-II}& 418.97 (4.98\%)& 7.42 \\
\hline
 \text{NSGA-II}& 408.36 (2.32\%) & 7.44  \\
 \text{with penalty} &  &   \\
\hline
\text{PODR} & 431.27 ( 8.06\%) & 7.44 \\
\hline
\text{MOIA} & 408.94 (2.46\%) & 7.43 \\
\hline
\text{Proposed method} & 401.20 (0.52\%)  & 7.43 \\
\hline
\end{tabular}
\label{tab_objective_values}
\end{table}

 Table~\ref{tab_objective_values} shows the average energy costs and user dissatisfaction
over the observation period. Two scenarios were considered. 
The first scenario assumed complete knowledge about future renewable power, electricity prices, and power load, i.e., no prediction errors were involved.   
The positive percentages associated with the energy costs in the parentheses indicate the cost increases as compared with the proposed method.
While similar levels of dissatisfaction were given,
the proposed method yielded the lowest energy costs, followed by MOIA, NSGA-II with penalty, NSGA-II with dominance definition, and PODR in order.   

The second scenario in Table~\ref{tab_objective_values} examined the effect of prediction errors. The energy cost of the proposed method from the first scenario was used as the baseline, i.e., the performance without prediction errors.
When further uncertainty was induced by the prediction errors,
the proposed method was able to address the uncertainty.
The energy cost of the proposed method only increased by 0.52\%, outperforming existing multiobjective demand response management methods having   
 at least an increase of 2.3\%.

\section{Conclusion}\label{sec_Conclusion}

This study introduces a multiobjective model predictive control method for managing residential demand response where an energy storage system is involved. The method focuses on two key objectives: minimizing electricity costs and reducing user dissatisfaction. The approach effectively addresses uncertainties related to future renewable energy generation, electricity prices, and base load variations.
To optimize future energy management decisions, Laguerre functions were employed to model decision variables, and convex samplers were utilized to generate feasible decision vectors. This setup allowed for the design of a multiobjective evolutionary algorithm that efficiently explores only the feasible set.
Numerical analysis demonstrates that the proposed method outperforms comparable multiobjective demand response management techniques, both with and without data uncertainties due to prediction errors. Notably, while maintaining a similar level of user satisfaction, the proposed method exhibited only a marginal increase in energy costs under conditions of data uncertainty compared to existing approaches.

\section*{Declaration of generative AI and AI-assisted technologies in the writing process}
During the preparation of this work the author(s) used ChatGPT in order to improve the clarity and fluency of the English writing. After using this tool/service, the author(s) reviewed and edited the content as needed and take(s) full responsibility for the content of the publication.

\bibliographystyle{elsarticle-num}
\bibliography{mybib}

@book{coello2007evolutionary,
  title={Evolutionary algorithms for solving multi-objective problems},
  author={Coello, Carlos A Coello},
  year={2007},
  publisher={Springer}
}

@article{yu2022survey,
  title={A survey on knee-oriented multiobjective evolutionary optimization},
  author={Yu, Guo and Ma, Lianbo and Jin, Yaochu and Du, Wenli and Liu, Qiqi and Zhang, Hengmin},
  journal={IEEE transactions on evolutionary computation},
  volume={26},
  number={6},
  pages={1452--1472},
  year={2022},
  publisher={IEEE}
}

@article{yao2023multi,
  title={Multi-level model predictive control based multi-objective optimal energy management of integrated energy systems considering uncertainty},
  author={Yao, Leyi and Liu, Zeyuan and Chang, Weiguang and Yang, Qiang},
  journal={Renewable Energy},
  volume={212},
  pages={523--537},
  year={2023},
  publisher={Elsevier}
}

@article{ascione2016simulation,
  title={Simulation-based model predictive control by the multi-objective optimization of building energy performance and thermal comfort},
  author={Ascione, Fabrizio and Bianco, Nicola and De Stasio, Claudio and Mauro, Gerardo Maria and Vanoli, Giuseppe Peter},
  journal={Energy and Buildings},
  volume={111},
  pages={131--144},
  year={2016},
  publisher={Elsevier}
}

@article{hua2024multi,
  title={Multi-criteria evaluation of novel multi-objective model predictive control method for indoor thermal comfort},
  author={Hua, Pengmin and Wang, Haichao and Xie, Zichan and Lahdelma, Risto},
  journal={Energy},
  volume={289},
  pages={129883},
  year={2024},
  publisher={Elsevier}
}

@inproceedings{jin2017user,
  title={User-preference-driven model predictive control of residential building loads and battery storage for demand response},
  author={Jin, Xin and Baker, Kyri and Isley, Steven and Christensen, Dane},
  booktitle={2017 American Control Conference (ACC)},
  pages={4147--4152},
  year={2017},
  organization={IEEE}
}

@article{farrokhifar2021model,
  title={Model predictive control for demand side management in buildings: A survey},
  author={Farrokhifar, Meisam and Bahmani, Hamidreza and Faridpak, Behdad and Safari, Amin and Pozo, David and Aiello, Marco},
  journal={Sustainable Cities and Society},
  volume={75},
  pages={103381},
  year={2021},
  publisher={Elsevier}
}

@article{freire2020optimal,
  title={Optimal demand response management of a residential microgrid using model predictive control},
  author={Freire, Vlademir A and De Arruda, L{\'u}cia Val{\'e}ria Ramos and Bordons, Carlos and M{\'a}rquez, Juan Jos{\'e}},
  journal={IEEE Access},
  volume={8},
  pages={228264--228276},
  year={2020},
  publisher={IEEE}
}

@article{hua2024integrated,
  title={Integrated demand response method for heating multiple rooms based on fuzzy logic considering dynamic price},
  author={Hua, Pengmin and Wang, Haichao and Xie, Zichan and Lahdelma, Risto},
  journal={Energy},
  volume={307},
  pages={132577},
  year={2024},
  publisher={Elsevier}
}

@article{FENG2022112357,
title = {Data-driven personal thermal comfort prediction: A literature review},
journal = {Renewable and Sustainable Energy Reviews},
volume = {161},
pages = {112357},
year = {2022},
issn = {1364-0321},
author = {Yanxiao Feng and Shichao Liu and Julian Wang and Jing Yang and Ying-Ling Jao and Nan Wang}
}

@book{Saltelli2008,
  author    = {Andrea Saltelli and Marco Ratto and Terry Andres and Francesca Campolongo and Jessica Cariboni and Debora Gatelli and Michaela Saisana and Stefano Tarantola},
  title     = {Global Sensitivity Analysis: The Primer},
  year      = {2008},
  publisher = {John Wiley I\& Sons},
  doi       = {10.1002/9780470725184}
}

@book{Taguchi2007,
  author    = {Genichi Taguchi and Subir Chowdhury and Yuin Wu},
  title     = {Taguchi’s Quality Engineering Handbook},
  year      = {2007},
  publisher = {Wiley},
  address   = {Hoboken, NJ, USA}
}

@article{Siami2019,
  author    = {Mohammad Amirian Siami and Nasim Kehtarnavaz},
  title     = {A review on deep learning approaches for time series prediction},
  journal   = {Applied Intelligence},
  volume    = {49},
  pages     = {737--753},
  year      = {2019},
  doi       = {10.1007/s10489-018-1368-9}
}

@manual{EWOSASolarFeedInTariffs,
  title        = {Solar Feed-in Tariffs},
  organization = {Energy \& Water Ombudsman SA},
  year         = {2023},
  month        = {February},
  url          = {https://ewosa.com.au/assets/volumes/general-downloads/fact-sheets/Solar-feed-in-tariffs.pdf}
}

@book{Gautschi2004,
  author    = {Walter Gautschi},
  title     = {Orthogonal Polynomials: Computation and Approximation},
  year      = {2004},
  publisher = {Oxford University Press},
  address   = {Oxford, UK},
}

@article{pham2021deep,
  title={Deep learning for intelligent demand response and smart grids: A comprehensive survey},
  author={Pham, Quoc-Viet and Liyanage, Madhusanka and Deepa, N and VVSS, Mounik and Reddy, Shivani and Maddikunta, Praveen Kumar Reddy and Khare, Neelu and Gadekallu, Thippa Reddy and Hwang, Won-Joo and others},
  journal={arXiv preprint arXiv:2101.08013},
  year={2021}
}

@article{zhou2016smart,
  title={Smart home energy management systems: Concept, configurations, and scheduling strategies},
  author={Zhou, Bin and Li, Wentao and Chan, Ka Wing and Cao, Yijia and Kuang, Yonghong and Liu, Xi and Wang, Xiong},
  journal={Renewable and Sustainable Energy Reviews},
  volume={61},
  pages={30--40},
  year={2016},
  publisher={Elsevier}
}

@inproceedings{das2020model,
  title={A Model for Optimizing Cost of Energy and Dissatisfaction for Household Consumers in Smart Home},
  author={Das, Nilima R and Rai, Satyananda C and Nayak, Ajit},
  booktitle={Advances in Intelligent Computing and Communication: Proceedings of ICAC 2019},
  pages={523--531},
  year={2020},
  organization={Springer}
}

@article{mavromatidis2018review,
  title={A review of uncertainty characterisation approaches for the optimal design of distributed energy systems},
  author={Mavromatidis, Georgios and Orehounig, Kristina and Carmeliet, Jan},
  journal={Renewable and Sustainable Energy Reviews},
  volume={88},
  pages={258--277},
  year={2018},
  publisher={Elsevier}
}

@article{shirsat2021quantifying,
  title={Quantifying residential demand response potential using a mixture density recurrent neural network},
  author={Shirsat, Ashwin and Tang, Wenyuan},
  journal={International Journal of Electrical Power \& Energy Systems},
  volume={130},
  pages={106853},
  year={2021},
  publisher={Elsevier}
}

@article{he2012residential,
  title={Residential demand response behavior analysis based on Monte Carlo simulation: The case of Yinchuan in China},
  author={He, Yongxiu and Wang, Bing and Wang, Jianhui and Xiong, Wei and Xia, Tian},
  journal={Energy},
  volume={47},
  number={1},
  pages={230--236},
  year={2012},
  publisher={Elsevier}
}

@inproceedings{11Hegde,
  title={Optimal control of residential energy storage under price fluctuations},
  author={Peter {van de ven} and Hegde, Nidhi and Massouli{\'e}, Laurent and Salonidis, Theodoros},
  booktitle={Proc. IARIA Energy conference},
  address={Venice, Italy},
  month=may,
  year={2011},
}

@ARTICLE{15Paterakis,
  title={Optimal Household Appliances Scheduling Under Day-Ahead Pricing and Load-Shaping Demand Response Strategies},
  author={N. G. {Paterakis} and O. {Erdinç} and A. G. {Bakirtzis} and J. P. S. {Catalão}},
  journal={{IEEE} Transactions on Industrial Informatics},
  volume={11},
  pages={1509-1519},
  year={2015},
  month=jun,
  }

@ARTICLE{18Melhem,
author={F. Y. {Melhem} and O. {Grunder} and Z. {Hammoudan} and N. {Moubayed}},
journal={{IEEE} Transactions on Industry Applications},
title={Energy Management in Electrical Smart Grid Environment Using Robust Optimization Algorithm},
year={2018},
volume={54},
pages={2714-2726},
month=may,
}

@article{15Althaher,
  author={Althaher, Sereen and Mancarella, Pierluigi and Mutale, Joseph},
  journal={{IEEE} Transactions on Smart Grid},
  title={Automated demand response from home energy management system under dynamic pricing and power and comfort constraints},
  year={2015},
  volume={6},
  pages={1874-1883},
  month=jul,
}

@article{14Setlhaolo,
author ={Ditiro Setlhaolo and Xiaohua Xia and Jiangfeng Zhang},
journal = {Electric Power Systems Research},
title = {Optimal scheduling of household appliances for demand response},
year = {2014},
volume = {116},
pages = {24-28},
month=nov,
}

@ARTICLE{16Roh,
author={H. {Roh} and J. {Lee}},
journal={{IEEE} Transactions on Smart Grid},
title={Residential Demand Response Scheduling With Multiclass Appliances in the Smart Grid},
year={2016},
volume={7},
pages={94-104},
month=jan,
}

@ARTICLE{19Chang,
author={H.-H. {Chang} and W.-Y. {Chiu} and H. {Sun} and C.-M. {Chen}},
journal={{IEEE} Systems Journal},
title={User-Centric Multiobjective Approach to Privacy Preservation and Energy Cost Minimization in Smart Home},
year={2019},
volume={13},
pages={1030-1041},
month=mar,
}

@INPROCEEDINGS{10Pindoriya,
author={N. M. {Pindoriya} and S. N. {Singh} and K. Y. {Lee}},
booktitle={Proc. IEEE PES General Meeting},
title={A comprehensive survey on multi-objective evolutionary optimization in power system applications},
address={Providence, RI, USA},
month=jul,
year={2010},
pages={1-8},
}

@ARTICLE{17Soares,
author={A. {Soares} and A. {Gomes} and C. H. {Antunes} and C. {Oliveira}},
journal={{IEEE} Transactions on Industrial Informatics},
title={A Customized Evolutionary Algorithm for Multiobjective Management of Residential Energy Resources},
year={2017},
volume={13},
pages={492-501},
month=apr,
}

@article{14Soares,
  author = {Ana Soares and Carlos Henggeler Antunes and Carlos Oliveira and Alvaro Gomes},
  journal = {Energy},
  title = {A multi-objective genetic approach to domestic load scheduling in an energy management system},
  year = {2014},
  volume = {77},
  pages = {144-152},
  month=dec,
}

@ARTICLE{20Chiu,
  author={W.-Y. {Chiu} and J.-T. {Hsieh} and C.-M. {Chen}},
  journal={{IEEE} Transactions on Industrial Informatics},
  title={Pareto Optimal Demand Response Based on Energy Costs and Load Factor in Smart Grid},
  year={2020},
  volume={16},
  pages={1811-1822},
  month=sep,
}

@ARTICLE{15Chiu,
  author={W.-Y. {Chiu} and H. {Sun} and H. {Vincent Poor}},
  journal={{IEEE} Transactions on Smart Grid},
  title={A Multiobjective Approach to Multimicrogrid System Design},
  year={2015},
  volume={6},
  pages={2263-2272},
  month=sep,
}

@ARTICLE{09Pedrasa,
author={M. A. A. {Pedrasa} and T. D. {Spooner} and I. F. {MacGill}},
journal={{IEEE} Transactions on Power Systems},
title={Scheduling of Demand Side Resources Using Binary Particle Swarm Optimization},
year={2009},
volume={24},
pages={1173-1181},
month=aug,
}

@article{12Salinas,
  author={Salinas, Sergio and Li, Ming and Li, Pan},
  journal={{IEEE} Transactions on Smart Grid},
  title={Multi-objective optimal energy consumption scheduling in smart grids},
  year={2012},
  volume={4},
  pages={341-348},
  month=mar,
}

@ARTICLE{15Nguyen,
  author={H. T. {Nguyen} and D. T. {Nguyen} and L. B. {Le}},
  journal={{IEEE} Transactions on Smart Grid},
  title={Energy Management for Households With Solar Assisted Thermal Load Considering Renewable Energy and Price Uncertainty},
  year={2015},
  volume={6},
  pages={301-314},
  month=sep,
}

@ARTICLE{19Li,
author={S. {Li} and J. {Yang} and W. {Song} and A. {Chen}},
journal={{IEEE} Internet of Things Journal},
title={A Real-Time Electricity Scheduling for Residential Home Energy Management},
year={2019},
volume={6},
pages={2602-2611},
month=Apr,
}

@ARTICLE{19Luo,
  author={F. {Luo} and G. {Ranzi} and C. {Wan} and Z. {Xu} and Z. Y. {Dong}},
   journal={IEEE Transactions on Industrial Informatics},
  title={A Multistage Home Energy Management System With Residential Photovoltaic Penetration},
  year={2019},
  volume={15},
  pages={116-126},
  month=jan,
}

@ARTICLE{16Zhang,
  author={Y. {Zhang} and R. {Wang} and T. {Zhang} and Y. {Liu} and B. {Guo}},
  journal={IET Generation, Transmission and Distribution},
  title={Model predictive control-based operation management for a residential microgrid with considering forecast uncertainties and demand response strategies},
  year={2016},
  volume={10},
  pages={2367-2378},
  month=jul,
}

@article{17Manzoor,
  title={An intelligent hybrid heuristic scheme for smart metering based demand side management in smart homes},
  author={Manzoor, Awais and Javaid, Nadeem and Ullah, Ibrar and Abdul, Wadood and Almogren, Ahmad and Alamri, Atif},
  journal={Energies},
  year={2017},
  volume={10},
  pages={1258-1285},
  month=aug,
}

@article{16Ma,
title = {Residential power scheduling for demand response in smart grid},
author = {Kai Ma and Ting Yao and Jie Yang and Xinping Guan},
journal = {International Journal of Electrical Power \& Energy Systems},
year = {2016},
volume = {78},
pages = {320-325},
month=jun,
}

@book{09Wang,
  title={Model predictive control system design and implementation using MATLAB{\textregistered}},
  author={Wang, Liuping},
  year={2009},
  publisher={Springer Science \& Business Media}
}

\end{document}